\documentclass[aps,reprint]{revtex4-2}

\usepackage{graphicx}%
\usepackage{amsmath,amssymb,amsfonts}%
\usepackage{xcolor}%
\usepackage{siunitx}
\usepackage[T1]{fontenc} 
\usepackage{upgreek}

\begin{document}
\title{The interplay of ferroelectricity and magneto-transport in non-magnetic moir\'{e} superlattices}

\author{Siqi Jiang$^{1\dagger}$}
\author{Renjun Du$^{1\dagger}$}
\author{Jiawei Jiang$^{1,2\dagger}$}
\author{Gan Liu$^{1}$}
\author{Jiabei Huang$^{1}$}
\author{Yu Du$^{1}$}
\author{Yaqing Han$^{1}$}
\author{Jingkuan Xiao$^{1}$}
\author{Di Zhang$^{1}$}
\author{Fuzhuo Lian$^{1}$}
\author{Wanting Xu$^{1}$}
\author{Siqin Wang$^{1}$}
\author{Lei Qiao$^{3}$}
\author{Kenji Watanabe$^{4}$}
\author{Takashi Taniguchi$^{5}$}
\author{Xiaoxiang Xi$^{1}$}
\author{Wei Ren$^{3}$}
\author{Baigeng Wang$^{1,6}$}
\author{Alexander S. Mayorov$^{1\ast}$}
\author{Kai Chang$^{2}$}
\author{Hongxin Yang$^{2\ast}$}
\author{Lei Wang$^{1,6\ast}$}
\author{Geliang Yu$^{1,6\ast}$}

\affiliation{$^{1}$National Laboratory of Solid State Microstructures, Collaborative Innovation Center of Advanced Microstructures, School of Physics, Nanjing University, Nanjing 210093, China.}

\affiliation{$^{2}$Center for Quantum Matter, School of Physics, Zhejiang University, Hangzhou 310058, China.}

\affiliation{$^{3}$Department of Physics, Shanghai University, Shanghai 200444, China.}

\affiliation{$^{4}$Research Center for Electronic and Optical Materials, National Institute for Materials Science, 1-1 Namiki, Tsukuba 305-0044, Japan.}

\affiliation{$^{5}$Research Center for Materials Nanoarchitectonics, National Institute for Materials Science, 1-1 Namiki, Tsukuba 305-0044, Japan.}

\affiliation{$^{6}$Jiangsu Physical Science Research Center, Nanjing University, Nanjing 210093, China.}

\affiliation{$^{\dagger}$These authors contributed equally to this work.}
\affiliation{$^{\ast}$Corresponding authors. E-mails: mayorov@nju.edu.cn; hongxin.yang@zju.edu.cn; leiwang@nju.edu.cn; yugeliang@nju.edu.cn}


\begin{abstract}

\begin{center}
\textbf{Abstract}
\end{center}

The coupling of ferroelectricity and magnetic order provides rich tunability for engineering material properties and demonstrates great potential for uncovering novel quantum phenomena and multifunctional devices. 
Here, we report interfacial ferroelectricity in moir\'{e} superlattices constructed from graphene and hexagonal boron nitride. 
We observe ferroelectric polarization in an across-layer moir\'{e} superlattice with an intercalated layer, demonstrating a remnant polarization comparable to its non-intercalated counterpart. 
Remarkably, we reveal a magnetic-field enhancement of ferroelectric polarization that persists up to room temperature, showcasing an unconventional amplification of ferroelectricity in materials lacking magnetic elements. 
This phenomenon, consistent across devices with varying layer configurations, arises purely from electronic rather than ionic contributions. 
Furthermore, the ferroelectric polarization in turn modulates quantum transport characteristics, suppressing Shubnikov-de Haas oscillations and altering quantum Hall states in polarized phases.
This interplay between ferroelectricity and magneto-transport in non-magnetic materials is crucial for exploring magnetoelectric effects and advancing two-dimensional memory and logic applications.
\end{abstract}

\maketitle

\section*{Introduction}

Ferroelectricity plays a significant role in modulating diverse material properties, including modulation of charge carrier density, lattice structures, electronic band spectra, and electron-phonon scattering, making it a fascinating area of research for uncovering novel physical mechanisms and functionalities~\cite{wang2023Twodimensional}.
Recent developments in van der Waals materials have broadened the scope of ferroelectrics from traditional bulk and thin-film systems into the atomic limit.
Of particular interest is the investigation of ferroelectricity in van der Waals heterostructures, which can be induced through the proximity effect~\cite{dey2023Thermally}, strain~\cite{xu2024Strain}, and polarized domain sliding~\cite{zhang2022Dominolike, Atri2024Spontaneous}, engineered via stacking and twisting of individual layers~\cite{zheng2020Unconventional, novoselov20162Da, Zhang2024Electronic, niu2024Unveiling, niu2022Giant, yan2023Moire}.
These findings exceed the conventional understanding that ferroelectricity is limited in materials with a polar space group, suggesting that symmetry breaking and interlayer interactions can induce electrical polarization in otherwise non-polar materials~\cite{wang2023Twodimensional}.
This enables the integration of ferroelectricity with intrinsic electronic properties, paving the way for multifunctional device design.
For instance, in twisted graphene multilayers, interfacial ferroelectricity has been linked to quantum states, such as ferroelectric superconductivity~\cite{klein2023Electrical} and ferroelectric Chern insulators~\cite{chen2024Selective}.

A particularly intriguing aspect of ferroelectricity is its interplay with magnetic fields, governed by magnetoelectric (ME) coupling, strain, and material-specific interactions. 
In multiferroics, this coupling can lead to technologically relevant phenomena, such as electrical-field-controlled magnetism or magnetic-field-modulated polarization~\cite{Kimura2003Magnetic, Cheong2007Multiferroics}. 
The underlying mechanisms—such as spin-orbit coupling (SOC) in SOC-dominated systems~\cite{Stroppa2014Tunable}, magnetostriction in strained materials~\cite{Hao2024Magnetic}, and exchange striction in magnetic compounds ~\cite{Yahia2017Recognition, Lee2011Mechanism, Sannigrahi2015Exchange}—typically require the presence of magnetic elements or long-range magnetic order.
Recent studies, however, have revealed magnetic-field-tunable ferroelectricity even in non-magnetic van der Waals systems, including CuInP$_{2}$S$_{6}$/graphene heterostructures~\cite{dey2023Thermally} and multilayer graphene~\cite{datta2017Strong, pan2017Layer,chen2024Selective}. A theoretical work on mixed-stack tetralayer graphene suggests that the simultaneous breaking of spatial-inversion and time-reversal symmetry is crucial for such magnetic responses~\cite{sarsfield2024Magnetic}. Realizing and controlling ferroelectricity in non-magnetic materials under magnetic fields remains an emerging frontier, offering exciting opportunities to discover new quantum phenomena, yet this field demands further systematic exploration.

Here, we report the experimental observation of two-dimensional (2D) ferroelectricity with an unconventional enhancement under perpendicular magnetic fields in moir\'{e} superlattices with distinct layer-stacking configurations, such as an ABA-trilayer graphene/hexagonal boron nitride (TLG/hBN) moir\'{e} superlattice and an across-layer TLG/hBN superlattice separated by monolayer graphene (MLG). 
Notably, the ferroelectric polarization persists in the across-layer superlattice, extending the family of 2D ferroelectrics to intercalated moir\'{e} systems, where the intercalation layer can be tailored to engineer diverse electronic properties.
Most strikingly, we discover a magnetic-field-enhanced ferroelectric polarization, persisting up to room temperature, in these entirely non-magnetic systems. 
The induced polarization dramatically alters magneto-transport behavior, enabling suppression of Shubnikov-de Haas (SdH) oscillations and modulation of quantum Hall states. 
The interplay between ferroelectricity and magneto-transport provides new insights into the integration of ferroelectricity with quantum phenomena, which may enrich the design of future multifunctional devices.

\section*{Results}
\subsection*{Ferroelectricity in an across-layer moir\'{e} superlattice}

Moir\'{e} superlattices have been demonstrated to be an effective approach to creating novel 2D ferroelectrics, such as a variety of van der Waals multilayers using graphene and hBN as building blocks~\cite{zheng2020Unconventional, niu2022Giant, Zhang2024Electronic, waters2024origin, Niu2025Ferroelectricity, Lin2025Room}.
Engineering moir\'{e} superlattices via an intercalating layer shows great potential for tuning interfacial ferroelectricity~\cite{Chen2025Ferroelectricity, yang2025super}.
Here, we investigate an ABA-stacked  TLG/hBN moir\'{e} system decoupled by inserting a misaligned MLG lattice.
To better understand the underlying mechanism, we prepared three comparison sections with different configurations in the same device (D1), depicted in Fig.~\ref{fig:Fig1}a.
The three sections include a perfectly aligned TLG/hBN moir\'{e} superlattice (D1-A), a $15\si{\degree}$-rotated-MLG intercalated TLG/hBN moir\'{e} system (D1-B), and 
a MLG-intercalated TLG/hBN stack in proximity to a WSe$_{2}$ SOC substrate (D1-C).
The different layouts induce diverse band dispersion and thus distinguish transport characteristics (see Supplementary Fig.~S10).
For instance, part D1-A exhibits typical secondary Dirac points (SDPs, Supplementary Fig.~S10d), indicative of the formation of minibands in the moir\'{e} superlattice~\cite{Dean2013hofstadter, Ponomarenko2013Cloning}.
While, part D1-B presents two separated zero-density lines associated with TLG and MLG (Supplementary Fig.~S10e), signifying the decoupled low-energy band structures~\cite{rickhaus2020Electronic, mrenca-kolasinska2022Quantum}.
In part D1-C, the SOC can enable the spin-splitting in the band structures and generate rich spin textures~\cite{Zollner2022Proximity, Yang2024Twist}, allowing for further modification of the transport properties (Supplementary Fig.~S10f).

In the following, we probe the exhibition of ferroelectric features in device D1.  
Figure~\ref{fig:Fig1}b demonstrates discernible longitudinal resistance ($R_\mathrm{xx}$) hysteresis in part D1-A when sweeping the back-gate voltage ($V_\mathrm{b}$) forward (red curve, $-60\ \unit{V}$ to $60\ \unit{V}$) and backward (blue curve, $60\ \unit{V}$ to $-60\ \unit{V}$) for the zero top-gate voltage $V_\mathrm{t}=0\ \unit{V}$.
In the following text, $V_\mathrm{b}$ and $V_\mathrm{t}$ are normalized by the corresponding dielectric thicknesses $d_\mathrm{b}$ and $d_\mathrm{t}$, respectively.
The shift of the Dirac point indicates a change in charge carrier density ($n$), which is about $\Delta n=0.12\unit{\times 10^{12}\,cm^{-2}}$ corresponding to the remnant polarization  ($P_\mathrm{2D}=e\Delta n d$, with $e$ representing the elementary charge and $d=0.34\ \unit{nm}$ denoting the interlayer distance)~\cite{niu2022Giant} of $0.066\ \unit{pC\,m^{-1}}$.
In Fig.~\ref{fig:Fig1}c, we present the difference in Hall density ($\Delta n_\mathrm{H}=n_\mathrm{H}^\mathrm{b}-n_\mathrm{H}^\mathrm{f}$) between the forward ($n_\mathrm{H}^\mathrm{f}$) and backward ($n_\mathrm{H}^\mathrm{b}$)  scans, varying as a function of $V_\mathrm{t}/d_\mathrm{t}$ and $V_\mathrm{b}/d_\mathrm{b}$.
Here, $V_\mathrm{b}$ is swept continuously as the fast-scan axis while $V_\mathrm{t}$ is stepped after each $V_\mathrm{b}$ sweep.
A pronounced ferroelectric hysteresis with the specific screening of the back-gate voltage is observed in part D1-A due to the polarized stacking domains in the TLG/hBN moir\'{e} superlattice~\cite{Li2017binary, Yang2023Across}, which is consistent with previous studies on mono-, bi-, and tri-layer graphene~\cite{zheng2020Unconventional, niu2022Giant, Zhang2024Electronic, waters2024origin, Niu2025Ferroelectricity}.
However, we discover almost the same ferroelectric characteristics in parts D1-B (Fig.~\ref{fig:Fig1}e,f) and D1-C (Supplementary Fig.~S10f,i), albeit the TLG and hBN lattices, giving rise to the moir\'{e} pattern, are separated by a decoupled graphene layer.
Notably, the polarization in three different sections can be switched on with a small threshold voltage $V_\mathrm{b}\approx 0.6\ \unit{V}$ and undergo the same linear growth as enlarging the scan range by raising the initial back-gate voltage $V_\mathrm{b}^\mathrm{ini}$ (Fig.~\ref{fig:Fig1}d).

Interfacial ferroelectricity in 2D artificial moir\'{e} superlattices is driven by the sliding ferroelectricity mechanism, which is demonstrated by both theoretical models~\cite{Yang2023Across, Li2017binary,ruiz2023MixedStacking,yang2023Atypical} and microscopic experiments~\cite{Stern2021interfacial, Atri2024Spontaneous, ko2023Operando, Molino2023Ferroelectric, Zhang2023Visualizing}.
A moir\'{e} superlattice provides a network of different stacking domains.
In non-centrosymmetric domains, where both inversion and mirror symmetry are broken, charge imbalance between layers induces either up- or down-polarization.
In contrast, centrosymmetric domains remain non-polarized. 
Under a sufficiently strong electrical field, low-energy domains expand while high-energy domains shrink toward domain walls. 
By reversing the polarity of the electrical field, the domains with different polarization can be switched by interlayer sliding, enabling the occurrence of ferroelectricity.
In the TLG/hBN system, the ABCA and ABAC domains exhibit significantly lower energies than the ABAB domain, facilitating a ferroelectric transition between ABCA and ABAC in part D1-A (Fig.~\ref{fig:Fig1}g), with a switchable polarization of $0.4\ \unit{pC\,m^{-1}}$ (Fig.~\ref{fig:Fig1}h).
When a MLG lattice is inserted between TLG and hBN at a large twist angle, the interfaces become incommensurate, reducing interlayer friction and sliding energy barriers~\cite{yang2025super} (Fig.~\ref{fig:Fig1}i) compared to the TLG/hBN counterpart (Fig.~\ref{fig:Fig1}g).
The resulting low-energy domains retain non-centrosymmetric and inherit those atomic configurations from the TLG and hBN constituents (Fig.~\ref{fig:Fig1}i), enabling a new transition pathway, ABBCA \textrightarrow\ ABBAB \textrightarrow\ ABBAC.
The vertical polarization difference between ABBCA and ABBAC domains yields a switchable polarization of $0.04\ \unit{pC\,m^{-1}}$ (Fig.~\ref{fig:Fig1}j), significantly suppressed due to screening from the incommensurate MLG layer and consistent with theoretical predictions for the MoS$_{2}$/MLG/MoS$_{2}$ system~\cite{yang2025super}.
Whereas, we observe nearly the same transition threshold and vertical polarization for all three parts, indicating that the intercalated MLG layer does not affect the overall polarization, which distinguishes from the strong enhancement in the double-aligned hBN/MLG/hBN system~\cite{Lin2025Room} and the predicted suppression in the MoS$_{2}$/MLG/MoS$_{2}$ structure~\cite{yang2025super}, highlighting the complex nature of ferroelectricity in intercalated moir\'{e} superlattices which requires further investigation.

\subsection*{Magnetic field-enhanced ferroelectricity}

In addition to electrical fields, magnetic fields can enable ferroelectric response in systems with magnetic elements or spin order, such as multiferroic materials~\cite{Cheong2007Multiferroics}.
Here, we investigate the magnetic control of ferroelectric polarization in van der Waals heterostructures based on non-magnetic graphene and hBN.
Figure~\ref{fig:Fig2}a,b illustrates $R_\mathrm{xx}(V_\mathrm{b}/d_\mathrm{b}, V_\mathrm{t}/d_\mathrm{t})$ in part D1-C at $B=13.5\ \unit{T}$ for the forward (red arrow) and backward (blue arrow) sweeps, respectively.
We notice that the charge neutrality points (CNPs) form a high resistance line and start to bend when $V_\mathrm{b}/d_\mathrm{b}>0$.
The difference of the CNPs for the forward and backward sweeps is presented in Fig.~\ref{fig:Fig2}c, where $\Delta n_\mathrm{H}$ varies as a function of displacement fields $D$ and $n$.
We notice that $\Delta n_\mathrm{H}$ changes sign when $D$ is tuned from positive to negative, indicating a flip in electrical polarization.
The corresponding $P_\mathrm{2D}$, displayed in Fig.~\ref{fig:Fig2}d, exhibits strong asymmetry with respect to $D$, yielding  a maximum of $0.19\ \unit{pC\,m^{-1}}$ for positive $D$, which is larger than that of $-0.066\ \unit{pC\,m^{-1}}$ at $B=0$.
Additionally, the resistance at the CNPs ($R_\mathrm{xx}^\mathrm{CNP}$) demonstrates a large shift in its maximum along the $D$ axis between the forward (red) and backward (blue) scans, and also reproduces the $D$-asymmetric behavior in Fig.~\ref{fig:Fig2}c,d.
To better visualize the behavior of magnetically enhanced electrical polarization, we present the $R_\mathrm{xx}(V_\mathrm{b}/d_\mathrm{b}, B)$ maps for the forward (Fig.~\ref{fig:Fig2}f) and backward (Fig.~\ref{fig:Fig2}g) sweeps at $V_\mathrm{t}/d_\mathrm{t}=-57\ \unit{mV\,nm^{-1}}$.
The positions of the CNPs in Fig.~\ref{fig:Fig2}f,g are extracted and displayed in Fig.~\ref{fig:Fig2}h.
We found that the CNPs for the forward sweep (red circles) are continuously shifted to the negative $V_\mathrm{b}/d_\mathrm{b}$ side as $B$ increases and become stable for $B>5.6\ \unit{T}$, while their backward counterparts (blue circles) 
show negligible change.
 A switch of the relative positions of the CNPs for the forward (red curve) and backward (blue curve) scans is observed with the growth of $B$, as shown in the linecuts at $B=0.2$ (Fig.~\ref{fig:Fig2}l), $1.7$ (Fig.~\ref{fig:Fig2}k), and $13.5\ \unit{T}$ (Fig.~\ref{fig:Fig2}j).
This switch  results in an increase of the polarization from $-0.05$ to $0.08\ \unit{pC\,m^{-1}}$ for $V_\mathrm{t}/d_\mathrm{t}=-57\ \unit{mV\,nm^{-1}}$ (Fig.~\ref{fig:Fig2}i).
We also found that $R_\mathrm{xx}^\mathrm{CNP}$ for the forward and backward sweeps increases asymmetrically with $B$ (Fig.~\ref{fig:Fig2}m), leading to a more pronounced resistance peak for the backward sweep. 
This $B$-enhanced polarization requires a high $V_\mathrm{b}$ over $\sim28\ \unit{V}$ to initialize the effect (Supplementary Fig.~S17), which is distinct from the zero-$B$ ferroelectric behavior that can be triggered with a small threshold voltage of about $0.6\ \unit{V}$.

Furthermore, we consider the magnetic effect on ferroelectric hysteresis in a different electronic system, the TLG/hBN superlattice (D1-A), in which the band structure is reconstructed by the moir\'{e} potential.
Figure~\ref{fig:Fig3}a presents the variation of $R_\mathrm{xx}$ with respect to $V_\mathrm{b}/d_\mathrm{b}$ for $B$ ranging from $0$ to $12\ \unit{T}$ at $T=60\ \unit{K}$.
Here, $R_\mathrm{xx}$ is normalized by the maximum in the forward and backward scans at each $B$.
The grey-shaded region indicates the enlarged hysteresis upon applying magnetic fields.
In Fig.~\ref{fig:Fig3}b, $P_\mathrm{2D}$ is shown as a function of $B$ for temperatures increasing from $20$ to $275\ \unit{K}$.
It is evident that $B$-enhanced ferroelectric polarization even persists up to room temperature.
However, the rate of the $B$-enhancement is intensely reduced as the temperature rises (inset in Fig.~\ref{fig:Fig3}b), especially for $T<120\ \unit{K}$.
Notably, the $B$-enhanced polarization emerges in both the TLG/hBN moir\'{e} superlattice and MLG-intercalated across-layer moir\'{e} superlattice.
We have also examined this peculiar magnetic response of ferroelectricity in other comparison devices with different structures, including a MLG/hBN superlattice (D2-A), decoupled MLG/TLG systems (D1-B, D2-B, D3), and a decoupled MLG/MLG stack (D2-C), and realized good reproducibility.
Therefore, this magnetic enhancement depends neither on a specific electronic system nor on the layer thickness.
Nevertheless, these investigated systems share a common feature that is the presence of a moir\'{e} superlattice, which provides the non-centrosymmetric domains for the appearance of ferroelectricity, albeit the moir\'{e} superlattice is sometimes decoupled by a middle layer.

The coupling between ferroelectricity and magnetic order, for instance, the ME effect,
typically occurs in materials with magnetic elements or spin order, where various degrees of freedom, such as electron charge, spin, orbital moments, and structure distortions, may interplay with each other~\cite{Cheong2007Multiferroics, Kimura2003Magnetic}.
The simultaneous breaking of spatial-inversion and time-reversal symmetry is crucial for the ME effect.
In our moir\'{e}-superlattice-based ferroelectrics, spatial-inversion symmetry is inherently broken, while time-reversal symmetry is induced by an external magnetic field.
The ME response generally consists of two contributions, ionic and electronic~\cite{Bousquet2011Unexpectedly, Dasa2019Designing}.
The ionic contribution stems from lattice distortions induced by $B$, while the electronic contribution arises from spin or orbital responses to $B$.
However, in our non-magnetic materials, ionic cores cannot be displaced by $B$, eliminating any lattice-mediated ME effects. 
Moreover, since we observe similarly enhanced ferroelectricity under $B$ in both SOC-included (D1-C) and SOC-free (D1-B) configurations, we infer that SOC-driven interfacial spin textures (\textit{e.g.}, skyrmions) have a negligible influence.
In addition, we performed density functional theory calculations to investigate the electronic polarization and structural distortions in each polar domain of the TLG/hBN moir\'{e} superlattice under magnetic fields (see details in Supplementary Discussion S12). 
The simulations confirm the lack of ionic contributions in our non-magnetic system, demonstrating that the ME effect is purely electronic—unlike conventional ME systems, where polarization is predominantly ionic~\cite{Bousquet2011Unexpectedly, Dasa2019Designing}.
In TLG/hBN systems, the electronic polarization shows a linear dependence on the out-of-plane $B$ in both ABCA and ABAC polar domains (Supplementary Fig.~S29). 
Notably, the $B$-enhanced polarization is significantly stronger in the ABAC domain, exceeding that in the ABCA domain by an order of magnitude.
This difference aligns with our experimental observation that the CNP position is strongly influenced by $B$ during the forward sweep but remains nearly unchanged in the backward scan (Fig.~\ref{fig:Fig2}f-h), a consequence of electrical-field-switchable polar domains.

Figure~\ref{fig:Fig3}c,d illustrates the temperature dependence of $R_\mathrm{xx}$ in part D1-A at $B=12\ \unit{T}$ for the forward and backward sweeps, respectively.
In both sweep directions, $R_\mathrm{xx}$ exhibits metallic behavior (\textit{i.e.}, decreasing resistance with lowering temperature), consistent with the zero-field case (Fig.~\ref{fig:Fig3}e).
In contrast, $R_\mathrm{xx}$ in part D1-B shows distinct features between two sweep directions: while $R_\mathrm{xx}$ in the forward sweep maintains metallic characteristics, $R_\mathrm{xx}$ in the backward sweep exhibits insulating behavior (increasing resistance with decreasing temperature), demonstrating a clear departure from the metallic case at $B=0\ \unit{T}$.
This directional asymmetry suggests that the domain motion during switching ferroelectric polarity induces distinct lattice configurations, leading to different electronic band structures for the two sweep directions.

\subsection*{Effects of ferroelectric polarization on magneto-transport}

Ferroelectric polarization can in turn affect magneto-transport, such as SdH oscillations and quantum Hall effects (QHEs). 
Figure~\ref{fig:Fig4}a,b presents a close inspection of SdH oscillations in device D3 (see the full-range maps in Supplementary Fig.~S26) for the forward and backward scans, respectively.
There are two sets of SdH oscillations belonging to MLG (pink-shaded regions) and TLG (fringes without shading), respectively.
The set for TLG is suppressed when the electrical polarization is on (P phase; labeled by the red bar), while the oscillations are sustained in the non-polarized region (N phase; denoted by the gray bar).
However,  the SdH oscillations for MLG (pink-shaded regions) persist regardless of whether the ferroelectric polarization is on or off.
The linecuts at $V_\mathrm{t}/d_\mathrm{t}=270\ \unit{mV\,nm^{-1}}$ in Fig.~\ref{fig:Fig4}a,b are demonstrated in Fig.~\ref{fig:Fig4}c, explicitly depicting the variation of SdH oscillations across the P and N phase boundaries.
The corresponding Hall conductance $G_\mathrm{xy}$ (Fig.~\ref{fig:Fig4}d) shows the effect of ferroelectric polarization on QHEs.
For the N phase (grey-shaded) in both forward and backward sweep directions, SdH oscillations and QHEs are both enabled and display a negligible difference between the two sweep directions.
Whereas, when the forward and backward sweeps are in the opposite phases, N and P, the SdH oscillations nearly vanish for the P phase but remain unchanged in the N phase.
Simultaneously, the filling factor of Hall plateaus in the P phase (blue curve in Fig.~\ref{fig:Fig4}d) may reduce from an integer to a random fraction value,  compared to the case in the N phase (red curve).
The difference in $G_\mathrm{xy}$ can be ascribed to the formation of bounded charge dipoles that reduce the number of mobile charges in the P phase.
Therefore, ferroelectricity may serve as an effective knob to control the on-off states of SdH oscillations and modify quantum Hall states.
At last, when both sweep directions are in the P phase (orange-shaded), the magneto-transport behavior is identical to the situation in the P phase.
There is an exception in the P phase that SdH oscillations and QHEs can be restored, though crossing with the MLG Landau levels, which do not seem to be disturbed by the ferroelectric polarization (pink-shaded regions in Fig.~\ref{fig:Fig4}e,f).

In ferroelectric polarized states, the suppression of SdH oscillations, indicating the anomalous vanishing of longitudinal signals, does not match with the stepped growth of the Hall conductance that suggests sustained transverse transport as well as the well-established Landau levels.
We have detected such anomalous suppression in all investigated devices with different layer configurations (Fig.~\ref{fig:Fig2}a,b, Supplementary Figs.~S13 and S21).
This behavior is sensitive to the scan range of $V_\mathrm{b}$ (Supplementary Fig.~S17) and arises when $V_\mathrm{b}$ is adequate to induce $B$-enhanced ferroelectric polarization, signified by the bending of the CNPs.
Moreover, we implemented a detailed examination of this peculiar magneto-transport in device D2-B (Supplementary Fig.~S27).
In the local measurements, we discover that in the P phase, the longitudinal resistances along two opposite edges of the current channel exhibit distinct behaviors, with SdH oscillations being suppressed at one edge but populated on the other side.
Moreover, when the magnetic field changes its sign, the situations along the two edges are exchanged, causing suppressed oscillations at the opposite edge.
This asymmetric behavior infers a non-uniform current distribution across the channel, which is further probed using nonlocal measurements (Supplementary Fig.~S28).

By passing the current along the transverse direction, the nonlocal resistance ($R_\mathrm{NL}=V_\mathrm{NL}/I_\mathrm{0}$; $V_\mathrm{NL}$, the nonlocal voltage; $I_\mathrm{0}$, the applied current) measured along the longitudinal edges on the left or right sides of the current path presents two key features.
i) On either side, SdH oscillations are asymmetric with respect to $B$.
For example, SdH oscillations are suppressed under negative $B$ (Supplementary Fig.~S28d) but sustained at positive $B$ (Supplementary Fig.~S28c) on the right side.
ii) For a fixed $B$, SdH oscillations display strong asymmetry between the left- and right-side edges of the current channel.
For instance, SdH oscillations are observed on the right side (Supplementary Fig.~S28c) but vanish on the left side (Supplementary Fig.~S28e) under positive $B$.
Reversing $B$ swaps the edges where SdH oscillations occur.
These findings align with local transport measurements, suggesting unidirectional propagation under magnetic fields.
 For a given $B$, only the edge on one side of the current channel permits SdH oscillations.
This unidirectional propagation resembles chiral transport in topological insulators under a magnetic field or magnetic topological insulators~\cite{Feng2019Tunable, Rosen2017Chiral, Zhu2025Direct}, where spatial-inversion and time-reversal symmetry are broken. 
In our system, ferroelectric polar domains break spatial-inversion symmetry, while the external magnetic field breaks time-reversal symmetry, potentially enabling analogous chiral transport. However, the precise link between magneto-transport and ferroelectric polarization remains unclear and demands further theoretical and microscopic investigation.

In conclusion, we demonstrate interfacial ferroelectricity in an across-layer moir\'{e} superlattice created by MLG intercalation, where the electrical polarization persists with undiminished strength despite the spatial separation of the moir\'{e} layers.
Remarkably, we discover a magnetic field enhancement of ferroelectric polarization that survives up to room temperature in these entirely non-magnetic systems. 
This effect, observed across devices with diverse layer configurations, originates purely from the electronic contributions, as evidenced by the exclusion of the ionic contributions. 
Furthermore, we demonstrate that ferroelectric polarization in turn modulates magneto-transport, suppressing SdH oscillations and altering quantum Hall states in polarized phases.
Nonlocal measurements reveal striking edge- and B-asymmetry of SdH oscillations in the ferroelectric polarized states, pointing to unidirectional propagation of currents.
These findings position graphene-hBN-based moir\'{e} superlattices as a unique non-magnetic platform for exploring magnetoelectric phenomena while opening new possibilities for 2D memory technologies.

\section*{Methods}
\subsection*{Device fabrication}
The van der Waals heterostructures were fabricated using the \textquotedblleft cut-and-stack\textquotedblright\ transfer method~\cite{Saito2020Independent}. 
The hBN and graphene flakes were exfoliated mechanically on Si substrates with $285\ \unit{nm}$ thermally oxidized SiO\textsubscript{2}.
We chose hBN flakes with a uniform thickness of $15\sim50\ \unit{nm}$ for dielectrics and encapsulation.
The graphene stripes with connected monolayer and trilayer parts were selected for further fabrication. 
The crystallographic axes of hBN were characterized with second harmonic generation (SHG) signals (Supplementary Fig.~S1a), and the crystal axes of graphene were determined using Raman spectroscopy~\cite{You2008Edge} measured along the zigzag and armchair edges (Supplementary Fig.~S1b).
The stacking order of trilayer graphene was characterized using Raman spectra (Supplementary Fig.~S2).
A microsurgical needle with a tip diameter of $1\ \unit{\upmu m}$ was employed to separate the monolayer and trilayer parts with the assistance of a high-precision manipulator.
The flakes are assembled from top to bottom on a hot plate at $90\si{\celsius}$ using a PDMS/PPC (Polydimethylsiloxane/polypropylene carbonate) stamp~\cite{Wang2013one, Kim2016van, cao2016superlattice}.
The top and bottom hBN flakes are $30\si{\degree}$ misaligned to ensure incommensurability in all samples.
In device D1, TLG is aligned with top hBN by $0\si{\degree}$, and MLG is rotated by $\sim 15\si{\degree}$ relative to TLG.
A flat part of the stack was selected to design the channel with a Hall-bar shape. 
The graphene layers and graphite gate were connected through metallic edge contacts (Ti/Au $5/60\ \unit{nm}$). 
The Hall-bar mesa was finally shaped using reactive ion etching (Supplementary Fig.~S3).

\subsection*{Measurements }

The devices were measured in an Oxford Instruments TeslatronPT cryogen-free system with a $14\ \unit{T}$ superconducting magnet and a base temperature of $\sim1.6\ \unit{K}$.
All electrical measurements were performed at $T=1.6\ \unit{K}$ unless indicated.
We have employed a standard low-frequency lock-in technique using SR830 lock-in amplifiers to apply an AC bias current of $10\sim100\ \unit{nA}$ at the excitation frequency of $13.333\sim 17.777\ \unit{Hz}$. To decrease the noise, we amplified the corresponding AC voltage $10^{3}$ times and current $10^{6}$ times by SR560 and SR570 preamplifiers, respectively. 
Keithley 2400/2614 source meters were used to apply voltages to the top and bottom gates. 
The charge carrier density ($n$) and displacement fields ($D$) can be controlled by tuning the voltages of the top gate ($V_\mathrm{t}$) and the Si back gate ($V_\mathrm{b}$).
The device is assumed to be a series of parallel-plate capacitors. 
The graphene multilayers are treated as a single charge layer coupled to the top and bottom gates with respective capacitance, $C_\mathrm{t}$ and $C_\mathrm{b}$.
We can calculate charge carrier density and displacement fields using the following relations: $n=(C_\mathrm{b}V_\mathrm{b}+C_\mathrm{t}V_\mathrm{t})/e$ and $D=(C_\mathrm{b}V_\mathrm{b}-C_\mathrm{t}V_\mathrm{t})/2\epsilon_\mathrm{0}$.
Here, $\epsilon_\mathrm{0}$ is the vacuum permittivity.

\subsection*{Determination of gate capacitance}
The capacitance is determined through the following procedures. First, we tuned top (bottom) gate voltages independently and measured Hall resistance at a small magnetic field ($0.1\ \unit{T}$) below the quantum Hall limit. 
The charge carrier density is calculated through the relation $R_\mathrm{xy}=\frac{B}{ne}$. 
The  corresponding gate capacitance is obtained by $C_\mathrm{i}=\frac{ne}{V_\mathrm{i}}$ ($\mathrm{i=t, b}$).
Then, to crosscheck the ratio of $C_\mathrm{t}$ to $C_\mathrm{b}$, we measured $R_\mathrm{xx}$ as a function of $V_\mathrm{t}$ and $V_\mathrm{b}$ at a constant magnetic field of $1\ \unit{T}$ and extracted the slope of the charge neutrality line, which is the ratio of the capacitance.
Finally, we fine-tune the capacitance by fitting the Landau Levels in the Landau fan diagram using $n=\frac{teB}{h}$.
Here, $h$ denotes Planck's constant, and $t$ is the filling factor of Landau levels.

\subsection*{Computational methods}
The first-principles calculations were performed with the projector augmented wave method as implemented in the Vienna ab initio simulation package~\cite{KRESSE1996Efficiency, Kresse1996Efficient}. The exchange-correlation effects were treated by the generalized gradient approximation in the Perdew-Burke-Ernzerhof form~\cite{Perdew1996Generalized}. Kohn-Sham single-particle wave functions were expanded in the plane wave basis set with a kinetic energy cutoff at $500\ \unit{eV}$. The energy and force convergence criteria were $10^{-7}\ \unit{eV}$ and $10^{-2}\ \si{eV\,\angstrom^{-1}}$, respectively. 
A $24\times24\times1$ $\mathrm{\Gamma}$-centered k-point mesh for the Brillouin zone integration was used for the heterostructures. A vacuum region of $15\ \si{\angstrom}$ was added in the out-of-plane direction to prevent the artificial coupling between the adjacent periodic images. The long-range correlation was included in evaluating van der Waals interaction by the optB86b method~\cite{Klime2011Van}. The dipole correction method~\cite{Neugebauer1992Adsorbate} was employed to evaluate the vertical polarization. The ferroelectric switching pathways were obtained by using the climbing image nudged elastic band method~\cite{Henkelman2000climbing}. The magnetic field was applied self-consistently by including SOC in the calculations~\cite{Bousquet2011Unexpectedly}. The electron polarization under finite magnetic fields was calculated using the Berry-phase approach~\cite{King-Smith1993Theory} with the positions of the ions frozen. Ionic relaxations in the presence of finite magnetic fields were performed to compute the change in the ionic polarization as a result of the magnetic fields by the Born effective charge approach~\cite{Gonze1997Dynamical}.

\section*{Data availability}
The data used in this study are available from the corresponding authors upon request.

\def\bibsection{

\section*{Acknowledgements}

The authors would like to thank Prof. Dr. Vladimir Fal'ko, Dr. Aitor Garcia-Ruiz, and Prof. Dr. Ming-Hao Liu for the fruitful discussion, the International Joint Lab of 2D Materials at Nanjing University for the support, and Yuanchen Co, Ltd (http://www.monosciences.com) for the High-Quality 2D Material Transfer System. 
G.Y. acknowledges the financial support from the National Key R\&D Program of China (Nos. 2024YFB3715400, 2022YFA1204700, and 2021YFA1400400), the National Natural Science Foundation of China (No. 11974169), the Natural Science Foundation of Jiangsu Province (No. BK20233001), and the support form Nanjing University International Collaboration Initiative. 
H.Y. acknowledges the National Key Research and Development Program of China (MOST) (No. 2022YFA1405102) for the financial support.
L.W. acknowledges the National Key Projects for Research and Development of China (Nos. 2022YFA1204700 and 2021YFA1400400), Natural Science Foundation of Jiangsu Province (Nos. BK20220066 and BK20233001). 
R.D. acknowledges the grant from the National Natural Science Foundation of China (No. 12004173). 
J.J. thanks the funding from the Postdoctoral Fellowship Program of CPSF (No. GZC20240694).

\section*{Author contributions}
G.Y. conceived the study; R.D. prepared the samples with the support of J.H., Y.H.,
D.Z., W.X., and S.W.; S.J. performed the transport measurements under the instruction of R.D., J.X., F.L., and A.S.M.; J.J., H.Y., W.R., and L.Q. worked out the theoretical model and implemented the numerical simulation with the support from B.W. and K.C.; G.L., Y.D., and X.X. conducted the SHG measurements; K.W. and T.T. grew the BN crystals; S.J. and R.D. analyzed the experimental data with the help of G.Y. and L.W.; R.D., S.J., and J.J. wrote the manuscript with input from G.Y., L.W., H.Y., and A.S.M.; all authors contributed to the discussion and commented on the manuscript.

\section*{Competing interests}
The authors declare no competing interests.


\begin{figure*}[h]
	\centering
	\includegraphics[scale=0.9]{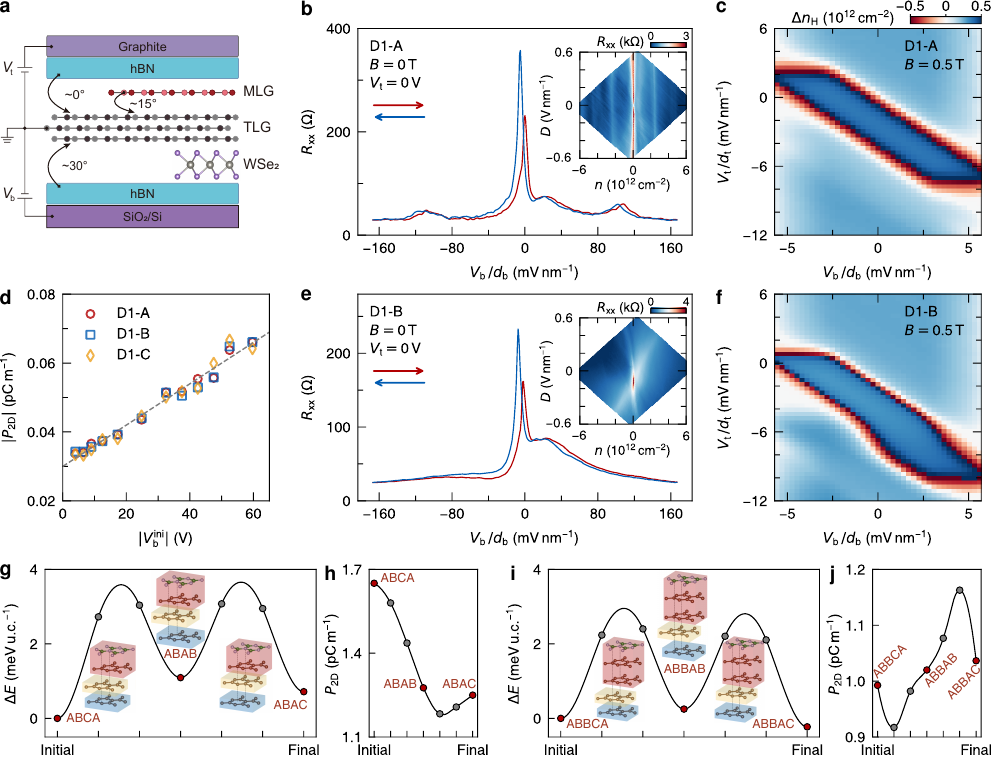}
	\caption{\textbf{Ferroelectric polarization in interfacial and across-layer moir\'{e} superlattices in the absence of magnetic fields.} 
			\textbf{a}, Schematics of the investigated device D1. There are three sections with distinct layer-stacking configurations, a $0\si{\degree}$-aligned ABA-stacked TLG/hBN moir\'{e} superlattice (D1-A), a $15\si{\degree}$-rotated-MLG intercalated TLG/hBN moir\'{e} system (D1-B), and a MLG-intercalated TLG/hBN stack in proximity to a WSe$_{2}$ substrate (D1-C).
			\textbf{b}, Longitudinal resistance $R_\mathrm{xx}$ varies with $V_\mathrm{b}/d_\mathrm{b}$ at $V_\mathrm{t}=0\ \unit{V}$ in part D1-A. 
			In this work, the back-gate ($V_\mathrm{b}$) and top-gate ($V_\mathrm{t}$) voltages  are normalized by the corresponding dielectric thicknesses $d_\mathrm{b}$ and $d_\mathrm{t}$, respectively.
			$R_\mathrm{xx}$ in the forward (red) and backward (blue) sweeps display evident resistance hysteresis, indicating ferroelectric polarization. The inset shows $R_\mathrm{xx}$ characterized with respect to charge carrier density $n$ and displacement fields $D$. The pronounced features of second Dirac points suggest the formation of a TLG/hBN moir\'{e} superlattice.
			\textbf{c}, The difference in Hall density ($\Delta n_\mathrm{H}=n_\mathrm{H}^\mathrm{b}-n_\mathrm{H}^\mathrm{f}$) between the forward ($n_\mathrm{H}^\mathrm{f}$) and backward ($n_\mathrm{H}^\mathrm{b}$)  scans (Supplementary Fig.~S11), varying as a function of $V_\mathrm{t}/d_\mathrm{t}$ and $V_\mathrm{b}/d_\mathrm{b}$ at $B=0.5\ \unit{T}$ for section D1-A.
			\textbf{d}, The absolute remnant polarization $|P_\mathrm{2D}|$ increases with enlarging the scan range. $V_\mathrm{b}^\mathrm{ini}$ denotes the initial back-gate voltage of each sweep. All three sections exhibit almost the same growth with $|V_\mathrm{b}^\mathrm{ini}|$.
			\textbf{e}, $R_\mathrm{xx}(V_\mathrm{b}/d_\mathrm{b})$ at $V_\mathrm{t}=0\ \unit{V}$ in part D1-B. 
			$R_\mathrm{xx}$ demonstrates a comparable hysteresis as in (\textbf{b}), indicating a non-diminished ferroelectric polarization in the across-layer moir\'{e} superlattice. The inset presents $R_\mathrm{xx}(n,D)$, illustrating the decoupling of MLG and TLG. 
			\textbf{f}, $\Delta n_\mathrm{H}(V_\mathrm{b}/d_\mathrm{b}, V_\mathrm{t}/d_\mathrm{t})$ at $B=0.5\ \unit{T}$ for section D1-B.
			\textbf{g},\textbf{h}, The energy per unit cell (u.c.) (\textbf{g}) and out-of-plane polarization  (\textbf{h}) for the transition path from ABCA to ABAC stacking configuration in the TLG/hBN system. 
			\textbf{i},\textbf{j},  The energy per u.c. (\textbf{i}) and out-of-plane polarization (\textbf{j}) for the transition path from ABBCA to ABBAC stacking configuration in the MLG-intercalated TLG/hBN system.}
	\label{fig:Fig1}
\end{figure*}

\begin{figure*}[hb]
	\centering
	\includegraphics[scale=0.9]{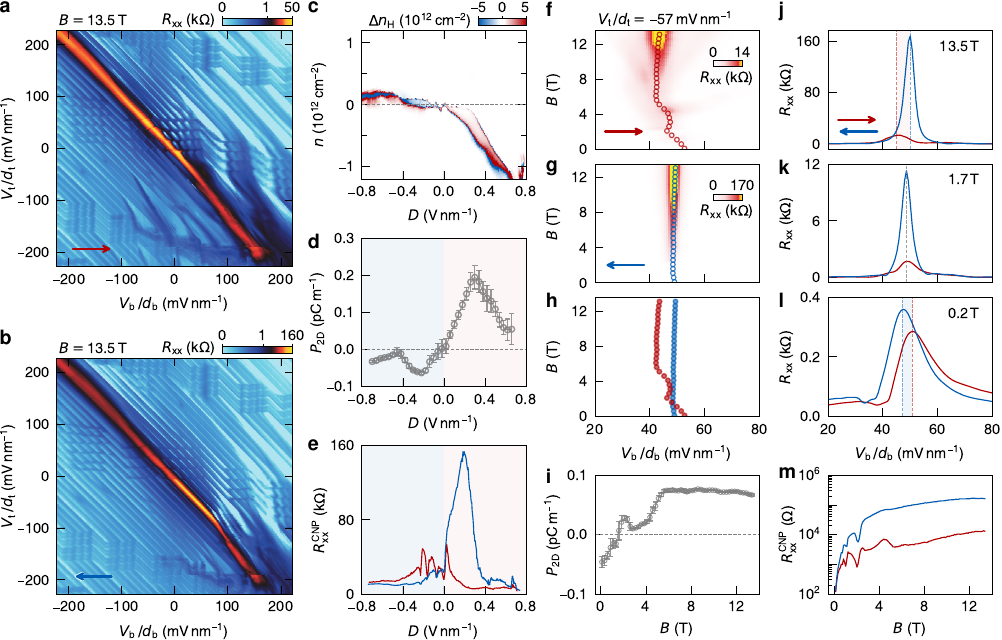}
	\caption{\textbf{Ferroelectric polarization enhanced by magnetic fields (D1-C).}
			\textbf{a},\textbf{b}, $R_\mathrm{xx}(V_\mathrm{b}/d_\mathrm{b}, V_\mathrm{t}/d_\mathrm{t})$ at $B=13.5\ \unit{T}$ for the forward (\textbf{a}) and backward (\textbf{b}) sweeps, respectively.
			\textbf{c}, $\Delta n_\mathrm{H}(D, n)$ presents the different positions of the CNPs for the two sweep directions.
			\textbf{d}, $P_\mathrm{2D}(D)$, corresponding to (\textbf{c}), is asymmetric with respect to $D$.
			\textbf{e}, The maximum of resistance at the CNPs ($R_\mathrm{xx}^\mathrm{CNP}$) shows a noticeable shift for the forward (red) and backward (blue) scans.
			\textbf{f},\textbf{g}, $R_\mathrm{xx}(V_\mathrm{b}/d_\mathrm{b}, B)$ at $V_\mathrm{t}/d_\mathrm{t}=-57\ \unit{mV\,nm^{-1}}$ for the forward (\textbf{f}) and backward (\textbf{g}) sweeps, respectively.
			\textbf{h}, The CNP positions extracted from (\textbf{f}) and (\textbf{g}) present a pronounced shift for the forward (red circles) and backward (blue circles) scans as $B$ increases.
			\textbf{i}, $P_\mathrm{2D}$, associated with (\textbf{h}), raises from $-0.05$ to $0.08\ \unit{pC\,m^{-1}}$ for $V_\mathrm{t}/d_\mathrm{t}=-57\ \unit{mV\,nm^{-1}}$.
			\textbf{j}-\textbf{l}, The evolution of the CNP positions between the two scans for $B=0.2$ (\textbf{l}), $1.7$ (\textbf{k}), and $13.5\ \unit{T}$ (\textbf{j}), respectively.
			\textbf{m}, $R_\mathrm{xx}^\mathrm{CNP}$ increases asymmetrically for the forward (red) and backward (blue) directions.}
	\label{fig:Fig2}
\end{figure*}

\begin{figure*}[h]
	\centering
	\includegraphics[scale=0.9]{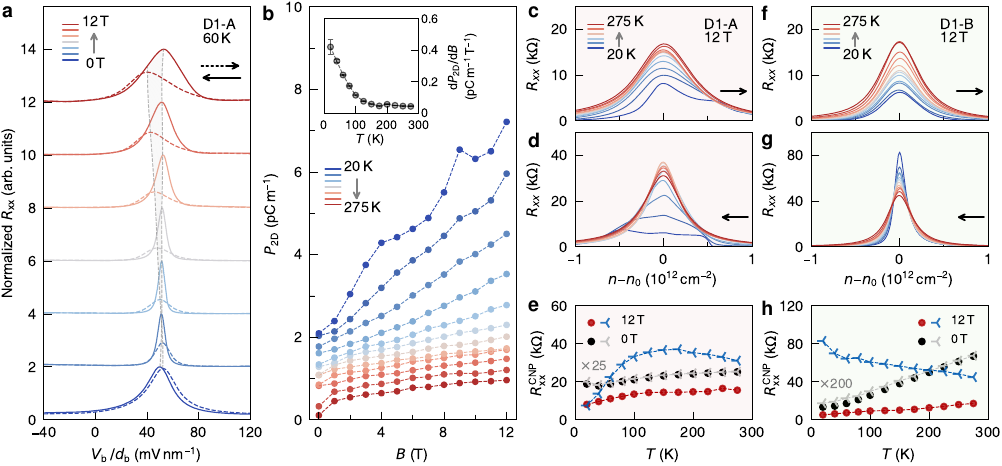}
	\caption{\textbf{Temperature dependence of $B$-enhanced ferroelectric polarization and evidence of ferroelectric domain change.}
			\textbf{a}, $R_\mathrm{xx}$ normalized by the maximum of both sweep directions at each $B$, demonstrates amplified hysteresis as $B$ increases from $0$ to $12\ \unit{T}$ for $V_\mathrm{t}/d_\mathrm{t}=-56\ \unit{mV\,nm^{-1}}$ at $T=60\ \unit{K}$ (D1-A).
			\textbf{b}, $P_\mathrm{2D}$ grows with $B$ for temperatures ranging from $20$ to $275\ \unit{K}$.
			Each curve is shifted for clarity.
			The inset displays the slope $\mathrm{d}P_\mathrm{2D}/\mathrm{d}B$ of each $P_\mathrm{2D}(B)$ curve, decreasing with increasing $T$ (D1-A).
			\textbf{c},\textbf{d}, The temperature dependence of $R_\mathrm{xx}$ at $B=12\ \unit{T}$ in part D1-A for the forward (\textbf{c}) and backward (\textbf{d}) sweeps, respectively.
			$n_\mathrm{0}$ denotes the charge carrier doping induced by raising the temperature.
			\textbf{e}, $R_\mathrm{xx}^\mathrm{CNP}$, corresponding to (\textbf{c}) and (\textbf{d}), for the forward (red dots) and backward (blue crosses) sweeps rises with $T$, demonstrating a metallic behavior as in the zero-$B$ case (amplified by $25$ times), shown as black dots and grey crosses for the forward and backward sweeps.
			\textbf{f},\textbf{g}, The temperature dependence of $R_\mathrm{xx}$ under $B=12\ \unit{T}$ in part D1-B for the forward (\textbf{f}) and backward (\textbf{g}) sweeps, respectively.
			\textbf{h}, $R_\mathrm{xx}^\mathrm{CNP}(T)$, corresponding to (\textbf{f}) and (\textbf{g}), for the forward (red dots) and backward (blue crosses) scans.
			In the forward sweep, $R_\mathrm{xx}^\mathrm{CNP}$ shows a metallic behavior, the same as that (black dots) in $B=0$ (enlarged by $200$ times). While in the backward sweep, it exhibits an insulating behavior, distinct from the zero-$B$ situation (grey crosses).
		}
	\label{fig:Fig3}
\end{figure*}

\begin{figure*}[h]
	\centering
	\includegraphics[scale=0.9]{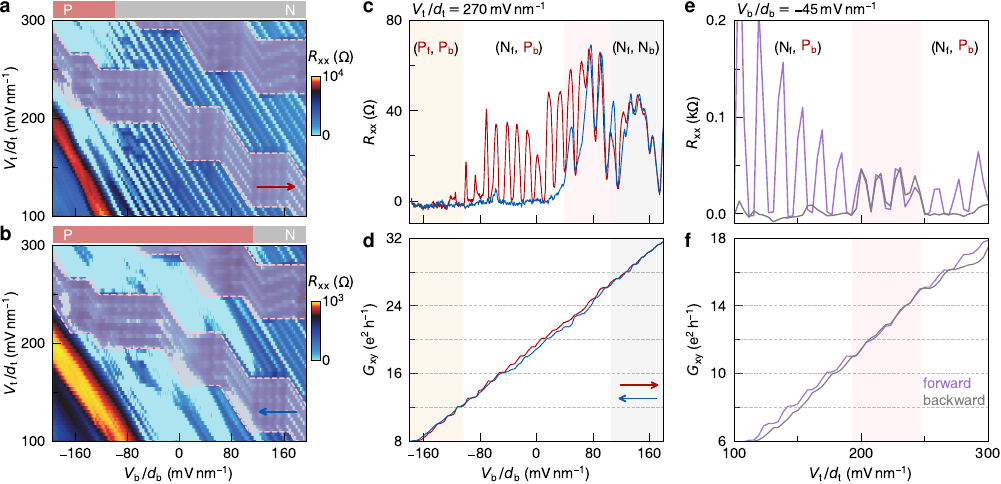}
	\caption{
			\textbf{The modulation of ferroelectric polarization on SdH oscillations and quantum Hall states (D3).}
			\textbf{a},\textbf{b}, $R_\mathrm{xx}(V_\mathrm{b}/d_\mathrm{b}, V_\mathrm{t}/d_\mathrm{t})$ at $B=13.6\ \unit{T}$ for the forward (\textbf{a}) and backward (\textbf{b}) sweeps, respectively, zoomed in from Supplementary Fig.~S26.
			The pink-shaded regions indicate the SdH oscillations arising from MLG, while the fringes without shading denote the SdH oscillations originating from TLG.
			The red and grey bars represent the polarized (P) and nonpolarized (N) phases.
			\textbf{c},\textbf{d}, $R_\mathrm{xx}$  (\textbf{c}) and $G_\mathrm{xy}$ (\textbf{d}) at $V_\mathrm{t}/d_\mathrm{t}=270\ \unit{mV\,nm^{-1}}$ for the forward (red) and backward (blue) scans, respectively.
			P\textsubscript{f} (N\textsubscript{f}) and P\textsubscript{b} (N\textsubscript{b}) are the polarized (nonpolarized) phases for forward and backward sweeps.
			\textbf{e},\textbf{f}, $R_\mathrm{xx}$ (\textbf{e}) and $G_\mathrm{xy}$ (\textbf{f}) at $V_\mathrm{b}/d_\mathrm{b}=-45\ \unit{mV\,nm^{-1}}$ for the forward (purple) and backward (grey) scans.}
	\label{fig:Fig4}
\end{figure*}


\section*{References}}
\begin{thebibliography}{60}%
	\makeatletter
	\providecommand \@ifxundefined [1]{%
		\@ifx{#1\undefined}
	}%
	\providecommand \@ifnum [1]{%
		\ifnum #1\expandafter \@firstoftwo
		\else \expandafter \@secondoftwo
		\fi
	}%
	\providecommand \@ifx [1]{%
		\ifx #1\expandafter \@firstoftwo
		\else \expandafter \@secondoftwo
		\fi
	}%
	\providecommand \natexlab [1]{#1}%
	\providecommand \enquote  [1]{``#1''}%
	\providecommand \bibnamefont  [1]{#1}%
	\providecommand \bibfnamefont [1]{#1}%
	\providecommand \citenamefont [1]{#1}%
	\providecommand \href@noop [0]{\@secondoftwo}%
	\providecommand \href [0]{\begingroup \@sanitize@url \@href}%
	\providecommand \@href[1]{\@@startlink{#1}\@@href}%
	\providecommand \@@href[1]{\endgroup#1\@@endlink}%
	\providecommand \@sanitize@url [0]{\catcode `\\12\catcode `\$12\catcode
		`\&12\catcode `\#12\catcode `\^12\catcode `\_12\catcode `\%12\relax}%
	\providecommand \@@startlink[1]{}%
	\providecommand \@@endlink[0]{}%
	\providecommand \url  [0]{\begingroup\@sanitize@url \@url }%
	\providecommand \@url [1]{\endgroup\@href {#1}{\urlprefix }}%
	\providecommand \urlprefix  [0]{URL }%
	\providecommand \Eprint [0]{\href }%
	\providecommand \doibase [0]{https://doi.org/}%
	\providecommand \selectlanguage [0]{\@gobble}%
	\providecommand \bibinfo  [0]{\@secondoftwo}%
	\providecommand \bibfield  [0]{\@secondoftwo}%
	\providecommand \translation [1]{[#1]}%
	\providecommand \BibitemOpen [0]{}%
	\providecommand \bibitemStop [0]{}%
	\providecommand \bibitemNoStop [0]{.\EOS\space}%
	\providecommand \EOS [0]{\spacefactor3000\relax}%
	\providecommand \BibitemShut  [1]{\csname bibitem#1\endcsname}%
	\let\auto@bib@innerbib\@empty
	\bibitem [{\citenamefont {Wang}\ \emph {et~al.}(2023)\citenamefont {Wang},
		\citenamefont {You}, \citenamefont {Cobden},\ and\ \citenamefont
		{Wang}}]{wang2023Twodimensional}%
	\BibitemOpen
	\bibfield  {author} {\bibinfo {author} {\bibfnamefont {C.}~\bibnamefont
			{Wang}}, \bibinfo {author} {\bibfnamefont {L.}~\bibnamefont {You}}, \bibinfo
		{author} {\bibfnamefont {D.}~\bibnamefont {Cobden}},\ and\ \bibinfo {author}
		{\bibfnamefont {J.}~\bibnamefont {Wang}},\ }\bibfield  {title} {\bibinfo
		{title} {Towards two-dimensional van der {{Waals}} ferroelectrics},\ }\href
	{https://doi.org/10.1038/s41563-022-01422-y} {\bibfield  {journal} {\bibinfo
			{journal} {Nat. Mater.}\ }\textbf {\bibinfo {volume} {22}},\ \bibinfo {pages}
		{542} (\bibinfo {year} {2023})}\BibitemShut {NoStop}%
	\bibitem [{\citenamefont {Dey}\ \emph {et~al.}(2023)\citenamefont {Dey},
		\citenamefont {Cottam}, \citenamefont {Makarovskiy}, \citenamefont {Yan},
		\citenamefont {Mi{\v s}eikis}, \citenamefont {Coletti}, \citenamefont
		{Kerfoot}, \citenamefont {Korolkov}, \citenamefont {Eaves}, \citenamefont
		{Linnartz}, \citenamefont {Kool}, \citenamefont {Wiedmann},\ and\
		\citenamefont {Patan{\`e}}}]{dey2023Thermally}%
	\BibitemOpen
	\bibfield  {author} {\bibinfo {author} {\bibfnamefont {A.}~\bibnamefont
			{Dey}}, \bibinfo {author} {\bibfnamefont {N.}~\bibnamefont {Cottam}},
		\bibinfo {author} {\bibfnamefont {O.}~\bibnamefont {Makarovskiy}}, \bibinfo
		{author} {\bibfnamefont {W.}~\bibnamefont {Yan}}, \bibinfo {author}
		{\bibfnamefont {V.}~\bibnamefont {Mi{\v s}eikis}}, \bibinfo {author}
		{\bibfnamefont {C.}~\bibnamefont {Coletti}}, \bibinfo {author} {\bibfnamefont
			{J.}~\bibnamefont {Kerfoot}}, \bibinfo {author} {\bibfnamefont
			{V.}~\bibnamefont {Korolkov}}, \bibinfo {author} {\bibfnamefont
			{L.}~\bibnamefont {Eaves}}, \bibinfo {author} {\bibfnamefont {J.~F.}\
			\bibnamefont {Linnartz}}, \bibinfo {author} {\bibfnamefont {A.}~\bibnamefont
			{Kool}}, \bibinfo {author} {\bibfnamefont {S.}~\bibnamefont {Wiedmann}},\
		and\ \bibinfo {author} {\bibfnamefont {A.}~\bibnamefont {Patan{\`e}}},\
	}\bibfield  {title} {\bibinfo {title} {Thermally stable quantum {Hall} effect
			in a gated ferroelectric-graphene heterostructure},\ }\href
	{https://doi.org/10.1038/s42005-023-01340-8} {\bibfield  {journal} {\bibinfo
			{journal} {Commun. Phys.}\ }\textbf {\bibinfo {volume} {6}},\ \bibinfo
		{pages} {216} (\bibinfo {year} {2023})}\BibitemShut {NoStop}%
	\bibitem [{\citenamefont {Xu}\ \emph {et~al.}(2024)\citenamefont {Xu},
		\citenamefont {Lomenzo}, \citenamefont {Kersch}, \citenamefont {Schenk},
		\citenamefont {Richter}, \citenamefont {Fancher}, \citenamefont {Starschich},
		\citenamefont {Berg}, \citenamefont {Reinig}, \citenamefont {Holsgrove},
		\citenamefont {Kiguchi}, \citenamefont {Mikolajick}, \citenamefont
		{Boettger},\ and\ \citenamefont {Schroeder}}]{xu2024Strain}%
	\BibitemOpen
	\bibfield  {author} {\bibinfo {author} {\bibfnamefont {B.}~\bibnamefont
			{Xu}}, \bibinfo {author} {\bibfnamefont {P.~D.}\ \bibnamefont {Lomenzo}},
		\bibinfo {author} {\bibfnamefont {A.}~\bibnamefont {Kersch}}, \bibinfo
		{author} {\bibfnamefont {T.}~\bibnamefont {Schenk}}, \bibinfo {author}
		{\bibfnamefont {C.}~\bibnamefont {Richter}}, \bibinfo {author} {\bibfnamefont
			{C.~M.}\ \bibnamefont {Fancher}}, \bibinfo {author} {\bibfnamefont
			{S.}~\bibnamefont {Starschich}}, \bibinfo {author} {\bibfnamefont
			{F.}~\bibnamefont {Berg}}, \bibinfo {author} {\bibfnamefont {P.}~\bibnamefont
			{Reinig}}, \bibinfo {author} {\bibfnamefont {K.~M.}\ \bibnamefont
			{Holsgrove}}, \bibinfo {author} {\bibfnamefont {T.}~\bibnamefont {Kiguchi}},
		\bibinfo {author} {\bibfnamefont {T.}~\bibnamefont {Mikolajick}}, \bibinfo
		{author} {\bibfnamefont {U.}~\bibnamefont {Boettger}},\ and\ \bibinfo
		{author} {\bibfnamefont {U.}~\bibnamefont {Schroeder}},\ }\bibfield  {title}
	{\bibinfo {title} {Strain as a global factor in stabilizing the ferroelectric
			properties of {{ZrO}}{\textsubscript{2}}},\ }\href
	{https://doi.org/10.1002/adfm.202311825} {\bibfield  {journal} {\bibinfo
			{journal} {Adv. Funct. Mater.}\ }\textbf {\bibinfo {volume} {34}},\ \bibinfo
		{pages} {2311825} (\bibinfo {year} {2024})}\BibitemShut {NoStop}%
	\bibitem [{\citenamefont {Zhang}\ \emph {et~al.}(2022)\citenamefont {Zhang},
		\citenamefont {Xu}, \citenamefont {Hou}, \citenamefont {Song}, \citenamefont
		{Ma}, \citenamefont {Gao}, \citenamefont {Zhu}, \citenamefont {Ma},
		\citenamefont {Liu}, \citenamefont {Feng},\ and\ \citenamefont
		{Li}}]{zhang2022Dominolike}%
	\BibitemOpen
	\bibfield  {author} {\bibinfo {author} {\bibfnamefont {S.}~\bibnamefont
			{Zhang}}, \bibinfo {author} {\bibfnamefont {Q.}~\bibnamefont {Xu}}, \bibinfo
		{author} {\bibfnamefont {Y.}~\bibnamefont {Hou}}, \bibinfo {author}
		{\bibfnamefont {A.}~\bibnamefont {Song}}, \bibinfo {author} {\bibfnamefont
			{Y.}~\bibnamefont {Ma}}, \bibinfo {author} {\bibfnamefont {L.}~\bibnamefont
			{Gao}}, \bibinfo {author} {\bibfnamefont {M.}~\bibnamefont {Zhu}}, \bibinfo
		{author} {\bibfnamefont {T.}~\bibnamefont {Ma}}, \bibinfo {author}
		{\bibfnamefont {L.}~\bibnamefont {Liu}}, \bibinfo {author} {\bibfnamefont
			{X.-Q.}\ \bibnamefont {Feng}},\ and\ \bibinfo {author} {\bibfnamefont
			{Q.}~\bibnamefont {Li}},\ }\bibfield  {title} {\bibinfo {title} {Domino-like
			stacking order switching in twisted monolayer--multilayer graphene},\ }\href
	{https://doi.org/10.1038/s41563-022-01232-2} {\bibfield  {journal} {\bibinfo
			{journal} {Nat. Mater.}\ }\textbf {\bibinfo {volume} {21}},\ \bibinfo {pages}
		{621} (\bibinfo {year} {2022})}\BibitemShut {NoStop}%
	\bibitem [{\citenamefont {Atri}\ \emph {et~al.}(2024)\citenamefont {Atri},
		\citenamefont {Cao}, \citenamefont {Alon}, \citenamefont {Roy}, \citenamefont
		{Stern}, \citenamefont {Falko}, \citenamefont {Goldstein}, \citenamefont
		{Kronik}, \citenamefont {Urbakh}, \citenamefont {Hod},\ and\ \citenamefont
		{Ben~Shalom}}]{Atri2024Spontaneous}%
	\BibitemOpen
	\bibfield  {author} {\bibinfo {author} {\bibfnamefont {S.~S.}\ \bibnamefont
			{Atri}}, \bibinfo {author} {\bibfnamefont {W.}~\bibnamefont {Cao}}, \bibinfo
		{author} {\bibfnamefont {B.}~\bibnamefont {Alon}}, \bibinfo {author}
		{\bibfnamefont {N.}~\bibnamefont {Roy}}, \bibinfo {author} {\bibfnamefont
			{M.~V.}\ \bibnamefont {Stern}}, \bibinfo {author} {\bibfnamefont
			{V.}~\bibnamefont {Falko}}, \bibinfo {author} {\bibfnamefont
			{M.}~\bibnamefont {Goldstein}}, \bibinfo {author} {\bibfnamefont
			{L.}~\bibnamefont {Kronik}}, \bibinfo {author} {\bibfnamefont
			{M.}~\bibnamefont {Urbakh}}, \bibinfo {author} {\bibfnamefont
			{O.}~\bibnamefont {Hod}},\ and\ \bibinfo {author} {\bibfnamefont
			{M.}~\bibnamefont {Ben~Shalom}},\ }\bibfield  {title} {\bibinfo {title}
		{Spontaneous electric polarization in graphene polytypes},\ }\href
	{https://doi.org/10.1002/apxr.202300095} {\bibfield  {journal} {\bibinfo
			{journal} {Adv. Phys. Res.}\ }\textbf {\bibinfo {volume} {3}},\ \bibinfo
		{pages} {2300095} (\bibinfo {year} {2024})}\BibitemShut {NoStop}%
	\bibitem [{\citenamefont {Zheng}\ \emph {et~al.}(2020)\citenamefont {Zheng},
		\citenamefont {Ma}, \citenamefont {Bi}, \citenamefont {De~La~Barrera},
		\citenamefont {Liu}, \citenamefont {Mao}, \citenamefont {Zhang},
		\citenamefont {Kiper}, \citenamefont {Watanabe}, \citenamefont {Taniguchi},
		\citenamefont {Kong}, \citenamefont {Tisdale}, \citenamefont {Ashoori},
		\citenamefont {Gedik}, \citenamefont {Fu}, \citenamefont {Xu},\ and\
		\citenamefont {{Jarillo-Herrero}}}]{zheng2020Unconventional}%
	\BibitemOpen
	\bibfield  {author} {\bibinfo {author} {\bibfnamefont {Z.}~\bibnamefont
			{Zheng}}, \bibinfo {author} {\bibfnamefont {Q.}~\bibnamefont {Ma}}, \bibinfo
		{author} {\bibfnamefont {Z.}~\bibnamefont {Bi}}, \bibinfo {author}
		{\bibfnamefont {S.}~\bibnamefont {De~La~Barrera}}, \bibinfo {author}
		{\bibfnamefont {M.-H.}\ \bibnamefont {Liu}}, \bibinfo {author} {\bibfnamefont
			{N.}~\bibnamefont {Mao}}, \bibinfo {author} {\bibfnamefont {Y.}~\bibnamefont
			{Zhang}}, \bibinfo {author} {\bibfnamefont {N.}~\bibnamefont {Kiper}},
		\bibinfo {author} {\bibfnamefont {K.}~\bibnamefont {Watanabe}}, \bibinfo
		{author} {\bibfnamefont {T.}~\bibnamefont {Taniguchi}}, \bibinfo {author}
		{\bibfnamefont {J.}~\bibnamefont {Kong}}, \bibinfo {author} {\bibfnamefont
			{W.~A.}\ \bibnamefont {Tisdale}}, \bibinfo {author} {\bibfnamefont
			{R.}~\bibnamefont {Ashoori}}, \bibinfo {author} {\bibfnamefont
			{N.}~\bibnamefont {Gedik}}, \bibinfo {author} {\bibfnamefont
			{L.}~\bibnamefont {Fu}}, \bibinfo {author} {\bibfnamefont {S.-Y.}\
			\bibnamefont {Xu}},\ and\ \bibinfo {author} {\bibfnamefont {P.}~\bibnamefont
			{{Jarillo-Herrero}}},\ }\bibfield  {title} {\bibinfo {title} {Unconventional
			ferroelectricity in moir{\'e} heterostructures},\ }\href
	{https://doi.org/10.1038/s41586-020-2970-9} {\bibfield  {journal} {\bibinfo
			{journal} {Nature}\ }\textbf {\bibinfo {volume} {588}},\ \bibinfo {pages}
		{71} (\bibinfo {year} {2020})}\BibitemShut {NoStop}%
	\bibitem [{\citenamefont {Novoselov}\ \emph {et~al.}(2016)\citenamefont
		{Novoselov}, \citenamefont {Mishchenko}, \citenamefont {Carvalho},\ and\
		\citenamefont {Castro~Neto}}]{novoselov20162Da}%
	\BibitemOpen
	\bibfield  {author} {\bibinfo {author} {\bibfnamefont {K.~S.}\ \bibnamefont
			{Novoselov}}, \bibinfo {author} {\bibfnamefont {A.}~\bibnamefont
			{Mishchenko}}, \bibinfo {author} {\bibfnamefont {A.}~\bibnamefont
			{Carvalho}},\ and\ \bibinfo {author} {\bibfnamefont {A.~H.}\ \bibnamefont
			{Castro~Neto}},\ }\bibfield  {title} {\bibinfo {title} {{{2D}} materials and
			van der {{Waals}} heterostructures},\ }\href
	{https://doi.org/10.1126/science.aac9439} {\bibfield  {journal} {\bibinfo
			{journal} {Science}\ }\textbf {\bibinfo {volume} {353}},\ \bibinfo {pages}
		{aac9439} (\bibinfo {year} {2016})}\BibitemShut {NoStop}%
	\bibitem [{\citenamefont {Zhang}\ \emph {et~al.}(2024)\citenamefont {Zhang},
		\citenamefont {Ding}, \citenamefont {Xiang}, \citenamefont {Liu},
		\citenamefont {Zhou}, \citenamefont {Wu}, \citenamefont {Xin}, \citenamefont
		{Watanabe}, \citenamefont {Taniguchi},\ and\ \citenamefont
		{Xu}}]{Zhang2024Electronic}%
	\BibitemOpen
	\bibfield  {author} {\bibinfo {author} {\bibfnamefont {L.}~\bibnamefont
			{Zhang}}, \bibinfo {author} {\bibfnamefont {J.}~\bibnamefont {Ding}},
		\bibinfo {author} {\bibfnamefont {H.}~\bibnamefont {Xiang}}, \bibinfo
		{author} {\bibfnamefont {N.}~\bibnamefont {Liu}}, \bibinfo {author}
		{\bibfnamefont {W.}~\bibnamefont {Zhou}}, \bibinfo {author} {\bibfnamefont
			{L.}~\bibnamefont {Wu}}, \bibinfo {author} {\bibfnamefont {N.}~\bibnamefont
			{Xin}}, \bibinfo {author} {\bibfnamefont {K.}~\bibnamefont {Watanabe}},
		\bibinfo {author} {\bibfnamefont {T.}~\bibnamefont {Taniguchi}},\ and\
		\bibinfo {author} {\bibfnamefont {S.}~\bibnamefont {Xu}},\ }\bibfield
	{title} {\bibinfo {title} {Electronic ferroelectricity in monolayer graphene
			moir\'{e} superlattices},\ }\href@noop {} {\bibfield  {journal} {\bibinfo
			{journal} {Nat. Commun.}\ }\textbf {\bibinfo {volume} {15}},\ \bibinfo
		{pages} {10905} (\bibinfo {year} {2024})}\BibitemShut {NoStop}%
	\bibitem [{\citenamefont {Niu}\ \emph {et~al.}(2025{\natexlab{a}})\citenamefont
		{Niu}, \citenamefont {Li}, \citenamefont {Han}, \citenamefont {Qu},
		\citenamefont {Liu}, \citenamefont {Wang}, \citenamefont {Han}, \citenamefont
		{Wang}, \citenamefont {Wu}, \citenamefont {Yang}, \citenamefont {Lv},
		\citenamefont {Yang}, \citenamefont {Watanabe}, \citenamefont {Taniguchi},
		\citenamefont {Liu}, \citenamefont {Mao}, \citenamefont {Shi}, \citenamefont
		{Che}, \citenamefont {Zhou}, \citenamefont {Xue}, \citenamefont {Wu},
		\citenamefont {Peng}, \citenamefont {Han}, \citenamefont {Gan},\ and\
		\citenamefont {Lu}}]{niu2024Unveiling}%
	\BibitemOpen
	\bibfield  {author} {\bibinfo {author} {\bibfnamefont {R.}~\bibnamefont
			{Niu}}, \bibinfo {author} {\bibfnamefont {Z.}~\bibnamefont {Li}}, \bibinfo
		{author} {\bibfnamefont {X.}~\bibnamefont {Han}}, \bibinfo {author}
		{\bibfnamefont {Z.}~\bibnamefont {Qu}}, \bibinfo {author} {\bibfnamefont
			{Q.}~\bibnamefont {Liu}}, \bibinfo {author} {\bibfnamefont {Z.}~\bibnamefont
			{Wang}}, \bibinfo {author} {\bibfnamefont {C.}~\bibnamefont {Han}}, \bibinfo
		{author} {\bibfnamefont {C.}~\bibnamefont {Wang}}, \bibinfo {author}
		{\bibfnamefont {Y.}~\bibnamefont {Wu}}, \bibinfo {author} {\bibfnamefont
			{C.}~\bibnamefont {Yang}}, \bibinfo {author} {\bibfnamefont {M.}~\bibnamefont
			{Lv}}, \bibinfo {author} {\bibfnamefont {K.}~\bibnamefont {Yang}}, \bibinfo
		{author} {\bibfnamefont {K.}~\bibnamefont {Watanabe}}, \bibinfo {author}
		{\bibfnamefont {T.}~\bibnamefont {Taniguchi}}, \bibinfo {author}
		{\bibfnamefont {K.}~\bibnamefont {Liu}}, \bibinfo {author} {\bibfnamefont
			{J.}~\bibnamefont {Mao}}, \bibinfo {author} {\bibfnamefont {W.}~\bibnamefont
			{Shi}}, \bibinfo {author} {\bibfnamefont {R.}~\bibnamefont {Che}}, \bibinfo
		{author} {\bibfnamefont {W.}~\bibnamefont {Zhou}}, \bibinfo {author}
		{\bibfnamefont {J.}~\bibnamefont {Xue}}, \bibinfo {author} {\bibfnamefont
			{M.}~\bibnamefont {Wu}}, \bibinfo {author} {\bibfnamefont {B.}~\bibnamefont
			{Peng}}, \bibinfo {author} {\bibfnamefont {Z.~V.}\ \bibnamefont {Han}},
		\bibinfo {author} {\bibfnamefont {Z.}~\bibnamefont {Gan}},\ and\ \bibinfo
		{author} {\bibfnamefont {J.}~\bibnamefont {Lu}},\ }\bibfield  {title}
	{\bibinfo {title} {Ferroelectricity with concomitant {C}oulomb screening in
			van der {W}aals heterostructures},\ }\href@noop {} {\bibfield  {journal}
		{\bibinfo  {journal} {Nat. Nanotechnol.}\ }\textbf {\bibinfo {volume} {20}},\
		\bibinfo {pages} {346} (\bibinfo {year} {2025}{\natexlab{a}})}\BibitemShut
	{NoStop}%
	\bibitem [{\citenamefont {Niu}\ \emph {et~al.}(2022)\citenamefont {Niu},
		\citenamefont {Li}, \citenamefont {Han}, \citenamefont {Qu}, \citenamefont
		{Ding}, \citenamefont {Wang}, \citenamefont {Liu}, \citenamefont {Liu},
		\citenamefont {Han}, \citenamefont {Watanabe}, \citenamefont {Taniguchi},
		\citenamefont {Wu}, \citenamefont {Ren}, \citenamefont {Wang}, \citenamefont
		{Hong}, \citenamefont {Mao}, \citenamefont {Han}, \citenamefont {Liu},
		\citenamefont {Gan},\ and\ \citenamefont {Lu}}]{niu2022Giant}%
	\BibitemOpen
	\bibfield  {author} {\bibinfo {author} {\bibfnamefont {R.}~\bibnamefont
			{Niu}}, \bibinfo {author} {\bibfnamefont {Z.}~\bibnamefont {Li}}, \bibinfo
		{author} {\bibfnamefont {X.}~\bibnamefont {Han}}, \bibinfo {author}
		{\bibfnamefont {Z.}~\bibnamefont {Qu}}, \bibinfo {author} {\bibfnamefont
			{D.}~\bibnamefont {Ding}}, \bibinfo {author} {\bibfnamefont {Z.}~\bibnamefont
			{Wang}}, \bibinfo {author} {\bibfnamefont {Q.}~\bibnamefont {Liu}}, \bibinfo
		{author} {\bibfnamefont {T.}~\bibnamefont {Liu}}, \bibinfo {author}
		{\bibfnamefont {C.}~\bibnamefont {Han}}, \bibinfo {author} {\bibfnamefont
			{K.}~\bibnamefont {Watanabe}}, \bibinfo {author} {\bibfnamefont
			{T.}~\bibnamefont {Taniguchi}}, \bibinfo {author} {\bibfnamefont
			{M.}~\bibnamefont {Wu}}, \bibinfo {author} {\bibfnamefont {Q.}~\bibnamefont
			{Ren}}, \bibinfo {author} {\bibfnamefont {X.}~\bibnamefont {Wang}}, \bibinfo
		{author} {\bibfnamefont {J.}~\bibnamefont {Hong}}, \bibinfo {author}
		{\bibfnamefont {J.}~\bibnamefont {Mao}}, \bibinfo {author} {\bibfnamefont
			{Z.}~\bibnamefont {Han}}, \bibinfo {author} {\bibfnamefont {K.}~\bibnamefont
			{Liu}}, \bibinfo {author} {\bibfnamefont {Z.}~\bibnamefont {Gan}},\ and\
		\bibinfo {author} {\bibfnamefont {J.}~\bibnamefont {Lu}},\ }\bibfield
	{title} {\bibinfo {title} {Giant ferroelectric polarization in a bilayer
			graphene heterostructure},\ }\href
	{https://doi.org/10.1038/s41467-022-34104-z} {\bibfield  {journal} {\bibinfo
			{journal} {Nat. Commun.}\ }\textbf {\bibinfo {volume} {13}},\ \bibinfo
		{pages} {6241} (\bibinfo {year} {2022})}\BibitemShut {NoStop}%
	\bibitem [{\citenamefont {Yan}\ \emph {et~al.}(2023)\citenamefont {Yan},
		\citenamefont {Zheng}, \citenamefont {Sangwan}, \citenamefont {Qian},
		\citenamefont {Wang}, \citenamefont {Liu}, \citenamefont {Watanabe},
		\citenamefont {Taniguchi}, \citenamefont {Xu}, \citenamefont
		{{Jarillo-Herrero}}, \citenamefont {Ma},\ and\ \citenamefont
		{Hersam}}]{yan2023Moire}%
	\BibitemOpen
	\bibfield  {author} {\bibinfo {author} {\bibfnamefont {X.}~\bibnamefont
			{Yan}}, \bibinfo {author} {\bibfnamefont {Z.}~\bibnamefont {Zheng}}, \bibinfo
		{author} {\bibfnamefont {V.~K.}\ \bibnamefont {Sangwan}}, \bibinfo {author}
		{\bibfnamefont {J.~H.}\ \bibnamefont {Qian}}, \bibinfo {author}
		{\bibfnamefont {X.}~\bibnamefont {Wang}}, \bibinfo {author} {\bibfnamefont
			{S.~E.}\ \bibnamefont {Liu}}, \bibinfo {author} {\bibfnamefont
			{K.}~\bibnamefont {Watanabe}}, \bibinfo {author} {\bibfnamefont
			{T.}~\bibnamefont {Taniguchi}}, \bibinfo {author} {\bibfnamefont {S.-Y.}\
			\bibnamefont {Xu}}, \bibinfo {author} {\bibfnamefont {P.}~\bibnamefont
			{{Jarillo-Herrero}}}, \bibinfo {author} {\bibfnamefont {Q.}~\bibnamefont
			{Ma}},\ and\ \bibinfo {author} {\bibfnamefont {M.~C.}\ \bibnamefont
			{Hersam}},\ }\bibfield  {title} {\bibinfo {title} {Moir{\'e} synaptic
			transistor with room-temperature neuromorphic functionality},\ }\href
	{https://doi.org/10.1038/s41586-023-06791-1} {\bibfield  {journal} {\bibinfo
			{journal} {Nature}\ }\textbf {\bibinfo {volume} {624}},\ \bibinfo {pages}
		{551} (\bibinfo {year} {2023})}\BibitemShut {NoStop}%
	\bibitem [{\citenamefont {Klein}\ \emph {et~al.}(2023)\citenamefont {Klein},
		\citenamefont {Xia}, \citenamefont {MacNeill}, \citenamefont {Watanabe},
		\citenamefont {Taniguchi},\ and\ \citenamefont
		{{Jarillo-Herrero}}}]{klein2023Electrical}%
	\BibitemOpen
	\bibfield  {author} {\bibinfo {author} {\bibfnamefont {D.~R.}\ \bibnamefont
			{Klein}}, \bibinfo {author} {\bibfnamefont {L.-Q.}\ \bibnamefont {Xia}},
		\bibinfo {author} {\bibfnamefont {D.}~\bibnamefont {MacNeill}}, \bibinfo
		{author} {\bibfnamefont {K.}~\bibnamefont {Watanabe}}, \bibinfo {author}
		{\bibfnamefont {T.}~\bibnamefont {Taniguchi}},\ and\ \bibinfo {author}
		{\bibfnamefont {P.}~\bibnamefont {{Jarillo-Herrero}}},\ }\bibfield  {title}
	{\bibinfo {title} {Electrical switching of a bistable moir{\'e}
			superconductor},\ }\href {https://doi.org/10.1038/s41565-022-01314-x}
	{\bibfield  {journal} {\bibinfo  {journal} {Nat. Nanotechnol.}\ }\textbf
		{\bibinfo {volume} {18}},\ \bibinfo {pages} {331} (\bibinfo {year}
		{2023})}\BibitemShut {NoStop}%
	\bibitem [{\citenamefont {Chen}\ \emph {et~al.}(2024)\citenamefont {Chen},
		\citenamefont {Xie}, \citenamefont {Cheng}, \citenamefont {Yang},
		\citenamefont {Li}, \citenamefont {Chen}, \citenamefont {Li}, \citenamefont
		{Xie}, \citenamefont {Watanabe}, \citenamefont {Taniguchi}, \citenamefont
		{He}, \citenamefont {Wu}, \citenamefont {Liang},\ and\ \citenamefont
		{Miao}}]{chen2024Selective}%
	\BibitemOpen
	\bibfield  {author} {\bibinfo {author} {\bibfnamefont {M.}~\bibnamefont
			{Chen}}, \bibinfo {author} {\bibfnamefont {Y.}~\bibnamefont {Xie}}, \bibinfo
		{author} {\bibfnamefont {B.}~\bibnamefont {Cheng}}, \bibinfo {author}
		{\bibfnamefont {Z.}~\bibnamefont {Yang}}, \bibinfo {author} {\bibfnamefont
			{X.-Z.}\ \bibnamefont {Li}}, \bibinfo {author} {\bibfnamefont
			{F.}~\bibnamefont {Chen}}, \bibinfo {author} {\bibfnamefont {Q.}~\bibnamefont
			{Li}}, \bibinfo {author} {\bibfnamefont {J.}~\bibnamefont {Xie}}, \bibinfo
		{author} {\bibfnamefont {K.}~\bibnamefont {Watanabe}}, \bibinfo {author}
		{\bibfnamefont {T.}~\bibnamefont {Taniguchi}}, \bibinfo {author}
		{\bibfnamefont {W.-Y.}\ \bibnamefont {He}}, \bibinfo {author} {\bibfnamefont
			{M.}~\bibnamefont {Wu}}, \bibinfo {author} {\bibfnamefont {S.-J.}\
			\bibnamefont {Liang}},\ and\ \bibinfo {author} {\bibfnamefont
			{F.}~\bibnamefont {Miao}},\ }\bibfield  {title} {\bibinfo {title} {Selective
			and quasi-continuous switching of ferroelectric {C}hern insulator devices for
			neuromorphic computing},\ }\href {https://doi.org/10.1038/s41565-024-01698-y}
	{\bibfield  {journal} {\bibinfo  {journal} {Nat. Nanotechnol.}\ }\textbf
		{\bibinfo {volume} {19}},\ \bibinfo {pages} {962} (\bibinfo {year}
		{2024})}\BibitemShut {NoStop}%
	\bibitem [{\citenamefont {Kimura}\ \emph {et~al.}(2003)\citenamefont {Kimura},
		\citenamefont {Goto}, \citenamefont {Shintani}, \citenamefont {Ishizaka},
		\citenamefont {Arima},\ and\ \citenamefont {Tokura}}]{Kimura2003Magnetic}%
	\BibitemOpen
	\bibfield  {author} {\bibinfo {author} {\bibfnamefont {T.}~\bibnamefont
			{Kimura}}, \bibinfo {author} {\bibfnamefont {T.}~\bibnamefont {Goto}},
		\bibinfo {author} {\bibfnamefont {H.}~\bibnamefont {Shintani}}, \bibinfo
		{author} {\bibfnamefont {K.}~\bibnamefont {Ishizaka}}, \bibinfo {author}
		{\bibfnamefont {T.}~\bibnamefont {Arima}},\ and\ \bibinfo {author}
		{\bibfnamefont {Y.}~\bibnamefont {Tokura}},\ }\bibfield  {title} {\bibinfo
		{title} {Magnetic control of ferroelectric polarization},\ }\href
	{https://doi.org/10.1038/nature02018} {\bibfield  {journal} {\bibinfo
			{journal} {Nature}\ }\textbf {\bibinfo {volume} {426}},\ \bibinfo {pages}
		{55} (\bibinfo {year} {2003})}\BibitemShut {NoStop}%
	\bibitem [{\citenamefont {Cheong}\ and\ \citenamefont
		{Mostovoy}(2007)}]{Cheong2007Multiferroics}%
	\BibitemOpen
	\bibfield  {author} {\bibinfo {author} {\bibfnamefont {S.-W.}\ \bibnamefont
			{Cheong}}\ and\ \bibinfo {author} {\bibfnamefont {M.}~\bibnamefont
			{Mostovoy}},\ }\bibfield  {title} {\bibinfo {title} {Multiferroics: a
			magnetic twist for ferroelectricity},\ }\href@noop {} {\bibfield  {journal}
		{\bibinfo  {journal} {Nat. Mater.}\ }\textbf {\bibinfo {volume} {6}},\
		\bibinfo {pages} {13} (\bibinfo {year} {2007})}\BibitemShut {NoStop}%
	\bibitem [{\citenamefont {Stroppa}\ \emph {et~al.}(2014)\citenamefont
		{Stroppa}, \citenamefont {Di~Sante}, \citenamefont {Barone}, \citenamefont
		{Bokdam}, \citenamefont {Kresse}, \citenamefont {Franchini}, \citenamefont
		{Whangbo},\ and\ \citenamefont {Picozzi}}]{Stroppa2014Tunable}%
	\BibitemOpen
	\bibfield  {author} {\bibinfo {author} {\bibfnamefont {A.}~\bibnamefont
			{Stroppa}}, \bibinfo {author} {\bibfnamefont {D.}~\bibnamefont {Di~Sante}},
		\bibinfo {author} {\bibfnamefont {P.}~\bibnamefont {Barone}}, \bibinfo
		{author} {\bibfnamefont {M.}~\bibnamefont {Bokdam}}, \bibinfo {author}
		{\bibfnamefont {G.}~\bibnamefont {Kresse}}, \bibinfo {author} {\bibfnamefont
			{C.}~\bibnamefont {Franchini}}, \bibinfo {author} {\bibfnamefont {M.-H.}\
			\bibnamefont {Whangbo}},\ and\ \bibinfo {author} {\bibfnamefont
			{S.}~\bibnamefont {Picozzi}},\ }\bibfield  {title} {\bibinfo {title} {Tunable
			ferroelectric polarization and its interplay with spin--orbit coupling in tin
			iodide perovskites},\ }\href@noop {} {\bibfield  {journal} {\bibinfo
			{journal} {Nat. Commun.}\ }\textbf {\bibinfo {volume} {5}},\ \bibinfo {pages}
		{5900} (\bibinfo {year} {2014})}\BibitemShut {NoStop}%
	\bibitem [{\citenamefont {Hao}\ \emph {et~al.}(2024)\citenamefont {Hao},
		\citenamefont {Chang}, \citenamefont {Chang}, \citenamefont {Liu},
		\citenamefont {Ho}, \citenamefont {Lu}, \citenamefont {Yang}, \citenamefont
		{Kawakami}, \citenamefont {Chen}, \citenamefont {Liu}, \citenamefont {Lin},
		\citenamefont {Lu}, \citenamefont {Lan},\ and\ \citenamefont
		{Yeh}}]{Hao2024Magnetic}%
	\BibitemOpen
	\bibfield  {author} {\bibinfo {author} {\bibfnamefont {D.}~\bibnamefont
			{Hao}}, \bibinfo {author} {\bibfnamefont {W.-H.}\ \bibnamefont {Chang}},
		\bibinfo {author} {\bibfnamefont {Y.-C.}\ \bibnamefont {Chang}}, \bibinfo
		{author} {\bibfnamefont {W.-T.}\ \bibnamefont {Liu}}, \bibinfo {author}
		{\bibfnamefont {S.-Z.}\ \bibnamefont {Ho}}, \bibinfo {author} {\bibfnamefont
			{C.-H.}\ \bibnamefont {Lu}}, \bibinfo {author} {\bibfnamefont {T.~H.}\
			\bibnamefont {Yang}}, \bibinfo {author} {\bibfnamefont {N.}~\bibnamefont
			{Kawakami}}, \bibinfo {author} {\bibfnamefont {Y.-C.}\ \bibnamefont {Chen}},
		\bibinfo {author} {\bibfnamefont {M.-H.}\ \bibnamefont {Liu}}, \bibinfo
		{author} {\bibfnamefont {C.-L.}\ \bibnamefont {Lin}}, \bibinfo {author}
		{\bibfnamefont {T.-H.}\ \bibnamefont {Lu}}, \bibinfo {author} {\bibfnamefont
			{Y.-W.}\ \bibnamefont {Lan}},\ and\ \bibinfo {author} {\bibfnamefont {N.-C.}\
			\bibnamefont {Yeh}},\ }\bibfield  {title} {\bibinfo {title} {Magnetic
			field-induced polar order in monolayer molybdenum disulfide transistors},\
	}\href {https://doi.org/https://doi.org/10.1002/adma.202411393} {\bibfield
		{journal} {\bibinfo  {journal} {Adv. Mater.}\ }\textbf {\bibinfo {volume}
			{36}},\ \bibinfo {pages} {2411393} (\bibinfo {year} {2024})}\BibitemShut
	{NoStop}%
	\bibitem [{\citenamefont {Yahia}\ \emph {et~al.}(2017)\citenamefont {Yahia},
		\citenamefont {Damay}, \citenamefont {Chattopadhyay}, \citenamefont
		{Bal\'edent}, \citenamefont {Peng}, \citenamefont {Elkaim}, \citenamefont
		{Whitaker}, \citenamefont {Greenblatt}, \citenamefont {Lepetit},\ and\
		\citenamefont {Foury-Leylekian}}]{Yahia2017Recognition}%
	\BibitemOpen
	\bibfield  {author} {\bibinfo {author} {\bibfnamefont {G.}~\bibnamefont
			{Yahia}}, \bibinfo {author} {\bibfnamefont {F.}~\bibnamefont {Damay}},
		\bibinfo {author} {\bibfnamefont {S.}~\bibnamefont {Chattopadhyay}}, \bibinfo
		{author} {\bibfnamefont {V.}~\bibnamefont {Bal\'edent}}, \bibinfo {author}
		{\bibfnamefont {W.}~\bibnamefont {Peng}}, \bibinfo {author} {\bibfnamefont
			{E.}~\bibnamefont {Elkaim}}, \bibinfo {author} {\bibfnamefont
			{M.}~\bibnamefont {Whitaker}}, \bibinfo {author} {\bibfnamefont
			{M.}~\bibnamefont {Greenblatt}}, \bibinfo {author} {\bibfnamefont {M.-B.}\
			\bibnamefont {Lepetit}},\ and\ \bibinfo {author} {\bibfnamefont
			{P.}~\bibnamefont {Foury-Leylekian}},\ }\bibfield  {title} {\bibinfo {title}
		{Recognition of exchange striction as the origin of magnetoelectric coupling
			in multiferroics},\ }\href {https://doi.org/10.1103/PhysRevB.95.184112}
	{\bibfield  {journal} {\bibinfo  {journal} {Phys. Rev. B}\ }\textbf {\bibinfo
			{volume} {95}},\ \bibinfo {pages} {184112} (\bibinfo {year}
		{2017})}\BibitemShut {NoStop}%
	\bibitem [{\citenamefont {Lee}\ \emph {et~al.}(2011)\citenamefont {Lee},
		\citenamefont {Choi}, \citenamefont {Ramazanoglu}, \citenamefont {Ratcliff},
		\citenamefont {Kiryukhin},\ and\ \citenamefont {Cheong}}]{Lee2011Mechanism}%
	\BibitemOpen
	\bibfield  {author} {\bibinfo {author} {\bibfnamefont {N.}~\bibnamefont
			{Lee}}, \bibinfo {author} {\bibfnamefont {Y.~J.}\ \bibnamefont {Choi}},
		\bibinfo {author} {\bibfnamefont {M.}~\bibnamefont {Ramazanoglu}}, \bibinfo
		{author} {\bibfnamefont {W.}~\bibnamefont {Ratcliff}}, \bibinfo {author}
		{\bibfnamefont {V.}~\bibnamefont {Kiryukhin}},\ and\ \bibinfo {author}
		{\bibfnamefont {S.-W.}\ \bibnamefont {Cheong}},\ }\bibfield  {title}
	{\bibinfo {title} {Mechanism of exchange striction of ferroelectricity in
			multiferroic orthorhombic {HoMnO$_{3}$} single crystals},\ }\href
	{https://doi.org/10.1103/PhysRevB.84.020101} {\bibfield  {journal} {\bibinfo
			{journal} {Phys. Rev. B}\ }\textbf {\bibinfo {volume} {84}},\ \bibinfo
		{pages} {020101} (\bibinfo {year} {2011})}\BibitemShut {NoStop}%
	\bibitem [{\citenamefont {Sannigrahi}\ \emph {et~al.}(2015)\citenamefont
		{Sannigrahi}, \citenamefont {Bhowal}, \citenamefont {Giri}, \citenamefont
		{Majumdar},\ and\ \citenamefont {Dasgupta}}]{Sannigrahi2015Exchange}%
	\BibitemOpen
	\bibfield  {author} {\bibinfo {author} {\bibfnamefont {J.}~\bibnamefont
			{Sannigrahi}}, \bibinfo {author} {\bibfnamefont {S.}~\bibnamefont {Bhowal}},
		\bibinfo {author} {\bibfnamefont {S.}~\bibnamefont {Giri}}, \bibinfo {author}
		{\bibfnamefont {S.}~\bibnamefont {Majumdar}},\ and\ \bibinfo {author}
		{\bibfnamefont {I.}~\bibnamefont {Dasgupta}},\ }\bibfield  {title} {\bibinfo
		{title} {Exchange-striction induced giant ferroelectric polarization in
			copper-based multiferroic material
			{$\ensuremath{\alpha}\ensuremath{-}{\mathrm{Cu}}_{2}{\mathrm{V}}_{2}{\mathrm{O}}_{7}$}},\
	}\href {https://doi.org/10.1103/PhysRevB.91.220407} {\bibfield  {journal}
		{\bibinfo  {journal} {Phys. Rev. B}\ }\textbf {\bibinfo {volume} {91}},\
		\bibinfo {pages} {220407} (\bibinfo {year} {2015})}\BibitemShut {NoStop}%
	\bibitem [{\citenamefont {Datta}\ \emph {et~al.}(2017)\citenamefont {Datta},
		\citenamefont {Dey}, \citenamefont {Samanta}, \citenamefont {Agarwal},
		\citenamefont {Borah}, \citenamefont {Watanabe}, \citenamefont {Taniguchi},
		\citenamefont {Sensarma},\ and\ \citenamefont {Deshmukh}}]{datta2017Strong}%
	\BibitemOpen
	\bibfield  {author} {\bibinfo {author} {\bibfnamefont {B.}~\bibnamefont
			{Datta}}, \bibinfo {author} {\bibfnamefont {S.}~\bibnamefont {Dey}}, \bibinfo
		{author} {\bibfnamefont {A.}~\bibnamefont {Samanta}}, \bibinfo {author}
		{\bibfnamefont {H.}~\bibnamefont {Agarwal}}, \bibinfo {author} {\bibfnamefont
			{A.}~\bibnamefont {Borah}}, \bibinfo {author} {\bibfnamefont
			{K.}~\bibnamefont {Watanabe}}, \bibinfo {author} {\bibfnamefont
			{T.}~\bibnamefont {Taniguchi}}, \bibinfo {author} {\bibfnamefont
			{R.}~\bibnamefont {Sensarma}},\ and\ \bibinfo {author} {\bibfnamefont
			{M.~M.}\ \bibnamefont {Deshmukh}},\ }\bibfield  {title} {\bibinfo {title}
		{Strong electronic interaction and multiple quantum {Hall} ferromagnetic
			phases in trilayer graphene},\ }\href {https://doi.org/10.1038/ncomms14518}
	{\bibfield  {journal} {\bibinfo  {journal} {Nat. Commun.}\ }\textbf {\bibinfo
			{volume} {8}},\ \bibinfo {pages} {14518} (\bibinfo {year}
		{2017})}\BibitemShut {NoStop}%
	\bibitem [{\citenamefont {Pan}\ \emph {et~al.}(2017)\citenamefont {Pan},
		\citenamefont {Wu}, \citenamefont {Cheng}, \citenamefont {Che}, \citenamefont
		{Taniguchi}, \citenamefont {Watanabe}, \citenamefont {Lau},\ and\
		\citenamefont {Bockrath}}]{pan2017Layer}%
	\BibitemOpen
	\bibfield  {author} {\bibinfo {author} {\bibfnamefont {C.}~\bibnamefont
			{Pan}}, \bibinfo {author} {\bibfnamefont {Y.}~\bibnamefont {Wu}}, \bibinfo
		{author} {\bibfnamefont {B.}~\bibnamefont {Cheng}}, \bibinfo {author}
		{\bibfnamefont {S.}~\bibnamefont {Che}}, \bibinfo {author} {\bibfnamefont
			{T.}~\bibnamefont {Taniguchi}}, \bibinfo {author} {\bibfnamefont
			{K.}~\bibnamefont {Watanabe}}, \bibinfo {author} {\bibfnamefont {C.~N.}\
			\bibnamefont {Lau}},\ and\ \bibinfo {author} {\bibfnamefont {M.}~\bibnamefont
			{Bockrath}},\ }\bibfield  {title} {\bibinfo {title} {Layer polarizability and
			easy-axis quantum {Hall} ferromagnetism in bilayer graphene},\ }\href
	{https://doi.org/10.1021/acs.nanolett.7b00197} {\bibfield  {journal}
		{\bibinfo  {journal} {Nano Lett.}\ }\textbf {\bibinfo {volume} {17}},\
		\bibinfo {pages} {3416} (\bibinfo {year} {2017})}\BibitemShut {NoStop}%
	\bibitem [{\citenamefont {Sarsfield}\ \emph {et~al.}(2024)\citenamefont
		{Sarsfield}, \citenamefont {Garcia-Ruiz},\ and\ \citenamefont
		{Fal'ko}}]{sarsfield2024Magnetic}%
	\BibitemOpen
	\bibfield  {author} {\bibinfo {author} {\bibfnamefont {P.~J.}\ \bibnamefont
			{Sarsfield}}, \bibinfo {author} {\bibfnamefont {A.}~\bibnamefont
			{Garcia-Ruiz}},\ and\ \bibinfo {author} {\bibfnamefont {V.~I.}\ \bibnamefont
			{Fal'ko}},\ }\bibfield  {title} {\bibinfo {title} {Substrate, temperature,
			and magnetic field dependence of electric polarization in mixed-stacking
			tetralayer graphenes},\ }\href
	{https://doi.org/10.1103/PhysRevResearch.6.043324} {\bibfield  {journal}
		{\bibinfo  {journal} {Phys. Rev. Res.}\ }\textbf {\bibinfo {volume} {6}},\
		\bibinfo {pages} {043324} (\bibinfo {year} {2024})}\BibitemShut {NoStop}%
	\bibitem [{\citenamefont {Waters}\ \emph {et~al.}(2024)\citenamefont {Waters},
		\citenamefont {Waleffe}, \citenamefont {Thompson}, \citenamefont
		{Arreguin-Martinez}, \citenamefont {Fonseca}, \citenamefont {Poirier},
		\citenamefont {Edgar}, \citenamefont {Watanabe}, \citenamefont {Taniguchi},
		\citenamefont {Xu}, \citenamefont {Cobden},\ and\ \citenamefont
		{Yankowitz}}]{waters2024origin}%
	\BibitemOpen
	\bibfield  {author} {\bibinfo {author} {\bibfnamefont {D.}~\bibnamefont
			{Waters}}, \bibinfo {author} {\bibfnamefont {D.}~\bibnamefont {Waleffe}},
		\bibinfo {author} {\bibfnamefont {E.}~\bibnamefont {Thompson}}, \bibinfo
		{author} {\bibfnamefont {E.}~\bibnamefont {Arreguin-Martinez}}, \bibinfo
		{author} {\bibfnamefont {J.}~\bibnamefont {Fonseca}}, \bibinfo {author}
		{\bibfnamefont {T.}~\bibnamefont {Poirier}}, \bibinfo {author} {\bibfnamefont
			{J.~H.}\ \bibnamefont {Edgar}}, \bibinfo {author} {\bibfnamefont
			{K.}~\bibnamefont {Watanabe}}, \bibinfo {author} {\bibfnamefont
			{T.}~\bibnamefont {Taniguchi}}, \bibinfo {author} {\bibfnamefont
			{X.}~\bibnamefont {Xu}}, \bibinfo {author} {\bibfnamefont {D.}~\bibnamefont
			{Cobden}},\ and\ \bibinfo {author} {\bibfnamefont {M.}~\bibnamefont
			{Yankowitz}},\ }\bibfield  {title} {\bibinfo {title} {On the origin of
			anomalous hysteresis in graphite/boron nitride transistors.},\ }\href@noop {}
	{\ ,\ \bibinfo {pages} {Preprint at https://arxiv.org/abs/2410.02699}
		(\bibinfo {year} {2024})},\ \Eprint {https://arxiv.org/abs/2410.02699}
	{arXiv:2410.02699 [cond-mat.mes-hall]} \BibitemShut {NoStop}%
	\bibitem [{\citenamefont {Niu}\ \emph {et~al.}(2025{\natexlab{b}})\citenamefont
		{Niu}, \citenamefont {Li}, \citenamefont {Han}, \citenamefont {Qu},
		\citenamefont {Liu}, \citenamefont {Wang}, \citenamefont {Han}, \citenamefont
		{Wang}, \citenamefont {Wu}, \citenamefont {Yang}, \citenamefont {Lv},
		\citenamefont {Yang}, \citenamefont {Watanabe}, \citenamefont {Taniguchi},
		\citenamefont {Liu}, \citenamefont {Mao}, \citenamefont {Shi}, \citenamefont
		{Che}, \citenamefont {Zhou}, \citenamefont {Xue}, \citenamefont {Wu},
		\citenamefont {Peng}, \citenamefont {Han}, \citenamefont {Gan},\ and\
		\citenamefont {Lu}}]{Niu2025Ferroelectricity}%
	\BibitemOpen
	\bibfield  {author} {\bibinfo {author} {\bibfnamefont {R.}~\bibnamefont
			{Niu}}, \bibinfo {author} {\bibfnamefont {Z.}~\bibnamefont {Li}}, \bibinfo
		{author} {\bibfnamefont {X.}~\bibnamefont {Han}}, \bibinfo {author}
		{\bibfnamefont {Z.}~\bibnamefont {Qu}}, \bibinfo {author} {\bibfnamefont
			{Q.}~\bibnamefont {Liu}}, \bibinfo {author} {\bibfnamefont {Z.}~\bibnamefont
			{Wang}}, \bibinfo {author} {\bibfnamefont {C.}~\bibnamefont {Han}}, \bibinfo
		{author} {\bibfnamefont {C.}~\bibnamefont {Wang}}, \bibinfo {author}
		{\bibfnamefont {Y.}~\bibnamefont {Wu}}, \bibinfo {author} {\bibfnamefont
			{C.}~\bibnamefont {Yang}}, \bibinfo {author} {\bibfnamefont {M.}~\bibnamefont
			{Lv}}, \bibinfo {author} {\bibfnamefont {K.}~\bibnamefont {Yang}}, \bibinfo
		{author} {\bibfnamefont {K.}~\bibnamefont {Watanabe}}, \bibinfo {author}
		{\bibfnamefont {T.}~\bibnamefont {Taniguchi}}, \bibinfo {author}
		{\bibfnamefont {K.}~\bibnamefont {Liu}}, \bibinfo {author} {\bibfnamefont
			{J.}~\bibnamefont {Mao}}, \bibinfo {author} {\bibfnamefont {W.}~\bibnamefont
			{Shi}}, \bibinfo {author} {\bibfnamefont {R.}~\bibnamefont {Che}}, \bibinfo
		{author} {\bibfnamefont {W.}~\bibnamefont {Zhou}}, \bibinfo {author}
		{\bibfnamefont {J.}~\bibnamefont {Xue}}, \bibinfo {author} {\bibfnamefont
			{M.}~\bibnamefont {Wu}}, \bibinfo {author} {\bibfnamefont {B.}~\bibnamefont
			{Peng}}, \bibinfo {author} {\bibfnamefont {Z.~V.}\ \bibnamefont {Han}},
		\bibinfo {author} {\bibfnamefont {Z.}~\bibnamefont {Gan}},\ and\ \bibinfo
		{author} {\bibfnamefont {J.}~\bibnamefont {Lu}},\ }\bibfield  {title}
	{\bibinfo {title} {Ferroelectricity with concomitant {C}oulomb screening in
			van der {W}aals heterostructures},\ }\href@noop {} {\bibfield  {journal}
		{\bibinfo  {journal} {Nat. Nanotechnol.}\ }\textbf {\bibinfo {volume} {20}},\
		\bibinfo {pages} {346} (\bibinfo {year} {2025}{\natexlab{b}})}\BibitemShut
	{NoStop}%
	\bibitem [{\citenamefont {Lin}\ \emph {et~al.}(2025)\citenamefont {Lin},
		\citenamefont {Xuan}, \citenamefont {Cao}, \citenamefont {Zhang},
		\citenamefont {Liu}, \citenamefont {Xue}, \citenamefont {Hang}, \citenamefont
		{Liu}, \citenamefont {Zhao}, \citenamefont {Gao}, \citenamefont {Guo},\ and\
		\citenamefont {Liu}}]{Lin2025Room}%
	\BibitemOpen
	\bibfield  {author} {\bibinfo {author} {\bibfnamefont {F.}~\bibnamefont
			{Lin}}, \bibinfo {author} {\bibfnamefont {X.}~\bibnamefont {Xuan}}, \bibinfo
		{author} {\bibfnamefont {Z.}~\bibnamefont {Cao}}, \bibinfo {author}
		{\bibfnamefont {Z.}~\bibnamefont {Zhang}}, \bibinfo {author} {\bibfnamefont
			{Y.}~\bibnamefont {Liu}}, \bibinfo {author} {\bibfnamefont {M.}~\bibnamefont
			{Xue}}, \bibinfo {author} {\bibfnamefont {Y.}~\bibnamefont {Hang}}, \bibinfo
		{author} {\bibfnamefont {X.}~\bibnamefont {Liu}}, \bibinfo {author}
		{\bibfnamefont {Y.}~\bibnamefont {Zhao}}, \bibinfo {author} {\bibfnamefont
			{L.}~\bibnamefont {Gao}}, \bibinfo {author} {\bibfnamefont {W.}~\bibnamefont
			{Guo}},\ and\ \bibinfo {author} {\bibfnamefont {Y.}~\bibnamefont {Liu}},\
	}\bibfield  {title} {\bibinfo {title} {Room temperature ferroelectricity in
			monolayer graphene sandwiched between hexagonal boron nitride},\ }\href@noop
	{} {\bibfield  {journal} {\bibinfo  {journal} {Nat. Commun.}\ }\textbf
		{\bibinfo {volume} {16}},\ \bibinfo {pages} {1189} (\bibinfo {year}
		{2025})}\BibitemShut {NoStop}%
	\bibitem [{\citenamefont {Chen}\ \emph {et~al.}(2025)\citenamefont {Chen},
		\citenamefont {Xuan}, \citenamefont {Guo},\ and\ \citenamefont
		{Zhang}}]{Chen2025Ferroelectricity}%
	\BibitemOpen
	\bibfield  {author} {\bibinfo {author} {\bibfnamefont {X.}~\bibnamefont
			{Chen}}, \bibinfo {author} {\bibfnamefont {X.}~\bibnamefont {Xuan}}, \bibinfo
		{author} {\bibfnamefont {W.}~\bibnamefont {Guo}},\ and\ \bibinfo {author}
		{\bibfnamefont {Z.}~\bibnamefont {Zhang}},\ }\bibfield  {title} {\bibinfo
		{title} {Ferroelectricity in van der {W}aals multilayers via interfacial
			polarization engineering},\ }\href@noop {} {\bibfield  {journal} {\bibinfo
			{journal} {npj 2D Mater. Appl.}\ }\textbf {\bibinfo {volume} {9}},\ \bibinfo
		{pages} {10} (\bibinfo {year} {2025})}\BibitemShut {NoStop}%
	\bibitem [{\citenamefont {Yang}\ and\ \citenamefont
		{Wu}(2025)}]{yang2025super}%
	\BibitemOpen
	\bibfield  {author} {\bibinfo {author} {\bibfnamefont {Z.}~\bibnamefont
			{Yang}}\ and\ \bibinfo {author} {\bibfnamefont {M.}~\bibnamefont {Wu}},\
	}\bibfield  {title} {\bibinfo {title} {Superlubric sliding
			ferroelectricity.},\ }\href@noop {} {\ ,\ \bibinfo {pages} {Preprint at
			https://arxiv.org/abs/2501.16118} (\bibinfo {year} {2025})},\ \Eprint
	{https://arxiv.org/abs/2501.16118} {arXiv:2501.16118 [cond-mat.mtrl-sci]}
	\BibitemShut {NoStop}%
	\bibitem [{\citenamefont {Dean}\ \emph {et~al.}(2013)\citenamefont {Dean},
		\citenamefont {Wang}, \citenamefont {Maher}, \citenamefont {Forsythe},
		\citenamefont {Ghahari}, \citenamefont {Gao}, \citenamefont {Katoch},
		\citenamefont {Ishigami}, \citenamefont {Moon}, \citenamefont {Koshino},
		\citenamefont {Taniguchi}, \citenamefont {Watanabe}, \citenamefont {Shepard},
		\citenamefont {Hone},\ and\ \citenamefont {Kim}}]{Dean2013hofstadter}%
	\BibitemOpen
	\bibfield  {author} {\bibinfo {author} {\bibfnamefont {C.~R.}\ \bibnamefont
			{Dean}}, \bibinfo {author} {\bibfnamefont {L.}~\bibnamefont {Wang}}, \bibinfo
		{author} {\bibfnamefont {P.}~\bibnamefont {Maher}}, \bibinfo {author}
		{\bibfnamefont {C.}~\bibnamefont {Forsythe}}, \bibinfo {author}
		{\bibfnamefont {F.}~\bibnamefont {Ghahari}}, \bibinfo {author} {\bibfnamefont
			{Y.}~\bibnamefont {Gao}}, \bibinfo {author} {\bibfnamefont {J.}~\bibnamefont
			{Katoch}}, \bibinfo {author} {\bibfnamefont {M.}~\bibnamefont {Ishigami}},
		\bibinfo {author} {\bibfnamefont {P.}~\bibnamefont {Moon}}, \bibinfo {author}
		{\bibfnamefont {M.}~\bibnamefont {Koshino}}, \bibinfo {author} {\bibfnamefont
			{T.}~\bibnamefont {Taniguchi}}, \bibinfo {author} {\bibfnamefont
			{K.}~\bibnamefont {Watanabe}}, \bibinfo {author} {\bibfnamefont {K.~L.}\
			\bibnamefont {Shepard}}, \bibinfo {author} {\bibfnamefont {J.}~\bibnamefont
			{Hone}},\ and\ \bibinfo {author} {\bibfnamefont {P.}~\bibnamefont {Kim}},\
	}\bibfield  {title} {\bibinfo {title} {Hofstadter's butterfly and the fractal
			quantum {H}all effect in moir{\'e} superlattices},\ }\href@noop {} {\bibfield
		{journal} {\bibinfo  {journal} {Nature}\ }\textbf {\bibinfo {volume}
			{497}},\ \bibinfo {pages} {598} (\bibinfo {year} {2013})}\BibitemShut
	{NoStop}%
	\bibitem [{\citenamefont {Ponomarenko}\ \emph {et~al.}(2013)\citenamefont
		{Ponomarenko}, \citenamefont {Gorbachev}, \citenamefont {Yu}, \citenamefont
		{Elias}, \citenamefont {Jalil}, \citenamefont {Patel}, \citenamefont
		{Mishchenko}, \citenamefont {Mayorov}, \citenamefont {Woods}, \citenamefont
		{Wallbank}, \citenamefont {Mucha-Kruczynski}, \citenamefont {Piot},
		\citenamefont {Potemski}, \citenamefont {Grigorieva}, \citenamefont
		{Novoselov}, \citenamefont {Guinea}, \citenamefont {Fal'ko},\ and\
		\citenamefont {Geim}}]{Ponomarenko2013Cloning}%
	\BibitemOpen
	\bibfield  {author} {\bibinfo {author} {\bibfnamefont {L.~A.}\ \bibnamefont
			{Ponomarenko}}, \bibinfo {author} {\bibfnamefont {R.~V.}\ \bibnamefont
			{Gorbachev}}, \bibinfo {author} {\bibfnamefont {G.~L.}\ \bibnamefont {Yu}},
		\bibinfo {author} {\bibfnamefont {D.~C.}\ \bibnamefont {Elias}}, \bibinfo
		{author} {\bibfnamefont {R.}~\bibnamefont {Jalil}}, \bibinfo {author}
		{\bibfnamefont {A.~A.}\ \bibnamefont {Patel}}, \bibinfo {author}
		{\bibfnamefont {A.}~\bibnamefont {Mishchenko}}, \bibinfo {author}
		{\bibfnamefont {A.~S.}\ \bibnamefont {Mayorov}}, \bibinfo {author}
		{\bibfnamefont {C.~R.}\ \bibnamefont {Woods}}, \bibinfo {author}
		{\bibfnamefont {J.~R.}\ \bibnamefont {Wallbank}}, \bibinfo {author}
		{\bibfnamefont {M.}~\bibnamefont {Mucha-Kruczynski}}, \bibinfo {author}
		{\bibfnamefont {B.~A.}\ \bibnamefont {Piot}}, \bibinfo {author}
		{\bibfnamefont {M.}~\bibnamefont {Potemski}}, \bibinfo {author}
		{\bibfnamefont {I.~V.}\ \bibnamefont {Grigorieva}}, \bibinfo {author}
		{\bibfnamefont {K.~S.}\ \bibnamefont {Novoselov}}, \bibinfo {author}
		{\bibfnamefont {F.}~\bibnamefont {Guinea}}, \bibinfo {author} {\bibfnamefont
			{V.~I.}\ \bibnamefont {Fal'ko}},\ and\ \bibinfo {author} {\bibfnamefont
			{A.~K.}\ \bibnamefont {Geim}},\ }\bibfield  {title} {\bibinfo {title}
		{Cloning of {D}irac fermions in graphene superlattices},\ }\href@noop {}
	{\bibfield  {journal} {\bibinfo  {journal} {Nature}\ }\textbf {\bibinfo
			{volume} {497}},\ \bibinfo {pages} {594} (\bibinfo {year}
		{2013})}\BibitemShut {NoStop}%
	\bibitem [{\citenamefont {Rickhaus}\ \emph {et~al.}(2020)\citenamefont
		{Rickhaus}, \citenamefont {Liu}, \citenamefont {Kurpas}, \citenamefont
		{Kurzmann}, \citenamefont {Lee}, \citenamefont {Overweg}, \citenamefont
		{Eich}, \citenamefont {Pisoni}, \citenamefont {Taniguchi}, \citenamefont
		{Watanabe}, \citenamefont {Richter}, \citenamefont {Ensslin},\ and\
		\citenamefont {Ihn}}]{rickhaus2020Electronic}%
	\BibitemOpen
	\bibfield  {author} {\bibinfo {author} {\bibfnamefont {P.}~\bibnamefont
			{Rickhaus}}, \bibinfo {author} {\bibfnamefont {M.-H.}\ \bibnamefont {Liu}},
		\bibinfo {author} {\bibfnamefont {M.}~\bibnamefont {Kurpas}}, \bibinfo
		{author} {\bibfnamefont {A.}~\bibnamefont {Kurzmann}}, \bibinfo {author}
		{\bibfnamefont {Y.}~\bibnamefont {Lee}}, \bibinfo {author} {\bibfnamefont
			{H.}~\bibnamefont {Overweg}}, \bibinfo {author} {\bibfnamefont
			{M.}~\bibnamefont {Eich}}, \bibinfo {author} {\bibfnamefont {R.}~\bibnamefont
			{Pisoni}}, \bibinfo {author} {\bibfnamefont {T.}~\bibnamefont {Taniguchi}},
		\bibinfo {author} {\bibfnamefont {K.}~\bibnamefont {Watanabe}}, \bibinfo
		{author} {\bibfnamefont {K.}~\bibnamefont {Richter}}, \bibinfo {author}
		{\bibfnamefont {K.}~\bibnamefont {Ensslin}},\ and\ \bibinfo {author}
		{\bibfnamefont {T.}~\bibnamefont {Ihn}},\ }\bibfield  {title} {\bibinfo
		{title} {The electronic thickness of graphene},\ }\href
	{https://doi.org/10.1126/sciadv.aay8409} {\bibfield  {journal} {\bibinfo
			{journal} {Sci. Adv.}\ }\textbf {\bibinfo {volume} {6}},\ \bibinfo {pages}
		{eaay8409} (\bibinfo {year} {2020})}\BibitemShut {NoStop}%
	\bibitem [{\citenamefont {{Mre{\'n}ca-Kolasi{\'n}ska}}\ \emph
		{et~al.}(2022)\citenamefont {{Mre{\'n}ca-Kolasi{\'n}ska}}, \citenamefont
		{Rickhaus}, \citenamefont {Zheng}, \citenamefont {Richter}, \citenamefont
		{Ihn}, \citenamefont {Ensslin},\ and\ \citenamefont
		{Liu}}]{mrenca-kolasinska2022Quantum}%
	\BibitemOpen
	\bibfield  {author} {\bibinfo {author} {\bibfnamefont {A.}~\bibnamefont
			{{Mre{\'n}ca-Kolasi{\'n}ska}}}, \bibinfo {author} {\bibfnamefont
			{P.}~\bibnamefont {Rickhaus}}, \bibinfo {author} {\bibfnamefont
			{G.}~\bibnamefont {Zheng}}, \bibinfo {author} {\bibfnamefont
			{K.}~\bibnamefont {Richter}}, \bibinfo {author} {\bibfnamefont
			{T.}~\bibnamefont {Ihn}}, \bibinfo {author} {\bibfnamefont {K.}~\bibnamefont
			{Ensslin}},\ and\ \bibinfo {author} {\bibfnamefont {M.-H.}\ \bibnamefont
			{Liu}},\ }\bibfield  {title} {\bibinfo {title} {Quantum capacitive coupling
			between large-angle twisted graphene layers},\ }\href
	{https://doi.org/10.1088/2053-1583/ac5536} {\bibfield  {journal} {\bibinfo
			{journal} {2D Mater.}\ }\textbf {\bibinfo {volume} {9}},\ \bibinfo {pages}
		{025013} (\bibinfo {year} {2022})}\BibitemShut {NoStop}%
	\bibitem [{\citenamefont {Zollner}\ \emph {et~al.}(2022)\citenamefont
		{Zollner}, \citenamefont {Gmitra},\ and\ \citenamefont
		{Fabian}}]{Zollner2022Proximity}%
	\BibitemOpen
	\bibfield  {author} {\bibinfo {author} {\bibfnamefont {K.}~\bibnamefont
			{Zollner}}, \bibinfo {author} {\bibfnamefont {M.}~\bibnamefont {Gmitra}},\
		and\ \bibinfo {author} {\bibfnamefont {J.}~\bibnamefont {Fabian}},\
	}\bibfield  {title} {\bibinfo {title} {Proximity spin-orbit and exchange
			coupling in {ABA} and {ABC} trilayer graphene van der {W}aals
			heterostructures},\ }\href {https://doi.org/10.1103/PhysRevB.105.115126}
	{\bibfield  {journal} {\bibinfo  {journal} {Phys. Rev. B}\ }\textbf {\bibinfo
			{volume} {105}},\ \bibinfo {pages} {115126} (\bibinfo {year}
		{2022})}\BibitemShut {NoStop}%
	\bibitem [{\citenamefont {Yang}\ \emph {et~al.}(2024)\citenamefont {Yang},
		\citenamefont {Mart{\'\i}n-Garc{\'\i}a}, \citenamefont {Kim{\'a}k},
		\citenamefont {Schmoranzerov{\'a}}, \citenamefont {Dolan}, \citenamefont
		{Chi}, \citenamefont {Gobbi}, \citenamefont {N{\v e}mec}, \citenamefont
		{Hueso},\ and\ \citenamefont {Casanova}}]{Yang2024Twist}%
	\BibitemOpen
	\bibfield  {author} {\bibinfo {author} {\bibfnamefont {H.}~\bibnamefont
			{Yang}}, \bibinfo {author} {\bibfnamefont {B.}~\bibnamefont
			{Mart{\'\i}n-Garc{\'\i}a}}, \bibinfo {author} {\bibfnamefont
			{J.}~\bibnamefont {Kim{\'a}k}}, \bibinfo {author} {\bibfnamefont
			{E.}~\bibnamefont {Schmoranzerov{\'a}}}, \bibinfo {author} {\bibfnamefont
			{E.}~\bibnamefont {Dolan}}, \bibinfo {author} {\bibfnamefont
			{Z.}~\bibnamefont {Chi}}, \bibinfo {author} {\bibfnamefont {M.}~\bibnamefont
			{Gobbi}}, \bibinfo {author} {\bibfnamefont {P.}~\bibnamefont {N{\v e}mec}},
		\bibinfo {author} {\bibfnamefont {L.~E.}\ \bibnamefont {Hueso}},\ and\
		\bibinfo {author} {\bibfnamefont {F.}~\bibnamefont {Casanova}},\ }\bibfield
	{title} {\bibinfo {title} {Twist-angle-tunable spin texture in
			{WS}e$_2$/graphene van der {W}aals heterostructures},\ }\href@noop {}
	{\bibfield  {journal} {\bibinfo  {journal} {Nat. Mater.}\ }\textbf {\bibinfo
			{volume} {23}},\ \bibinfo {pages} {1502} (\bibinfo {year}
		{2024})}\BibitemShut {NoStop}%
	\bibitem [{\citenamefont {Li}\ and\ \citenamefont {Wu}(2017)}]{Li2017binary}%
	\BibitemOpen
	\bibfield  {author} {\bibinfo {author} {\bibfnamefont {L.}~\bibnamefont
			{Li}}\ and\ \bibinfo {author} {\bibfnamefont {M.}~\bibnamefont {Wu}},\
	}\bibfield  {title} {\bibinfo {title} {Binary compound bilayer and multilayer
			with vertical polarizations: {T}wo-dimensional ferroelectrics, multiferroics,
			and nanogenerators},\ }\href {https://doi.org/10.1021/acsnano.7b02756}
	{\bibfield  {journal} {\bibinfo  {journal} {ACS Nano}\ }\textbf {\bibinfo
			{volume} {11}},\ \bibinfo {pages} {6382} (\bibinfo {year}
		{2017})}\BibitemShut {NoStop}%
	\bibitem [{\citenamefont {Yang}\ and\ \citenamefont
		{Wu}(2023)}]{Yang2023Across}%
	\BibitemOpen
	\bibfield  {author} {\bibinfo {author} {\bibfnamefont {L.}~\bibnamefont
			{Yang}}\ and\ \bibinfo {author} {\bibfnamefont {M.}~\bibnamefont {Wu}},\
	}\bibfield  {title} {\bibinfo {title} {Across-layer sliding ferroelectricity
			in {2D} heterolayers},\ }\href
	{https://doi.org/https://doi.org/10.1002/adfm.202301105} {\bibfield
		{journal} {\bibinfo  {journal} {Adv. Funct. Mater.}\ }\textbf {\bibinfo
			{volume} {33}},\ \bibinfo {pages} {2301105} (\bibinfo {year}
		{2023})}\BibitemShut {NoStop}%
	\bibitem [{\citenamefont {{Garcia-Ruiz}}\ \emph {et~al.}(2023)\citenamefont
		{{Garcia-Ruiz}}, \citenamefont {Enaldiev}, \citenamefont {McEllistrim},\ and\
		\citenamefont {Fal'ko}}]{ruiz2023MixedStacking}%
	\BibitemOpen
	\bibfield  {author} {\bibinfo {author} {\bibfnamefont {A.}~\bibnamefont
			{{Garcia-Ruiz}}}, \bibinfo {author} {\bibfnamefont {V.}~\bibnamefont
			{Enaldiev}}, \bibinfo {author} {\bibfnamefont {A.}~\bibnamefont
			{McEllistrim}},\ and\ \bibinfo {author} {\bibfnamefont {V.~I.}\ \bibnamefont
			{Fal'ko}},\ }\bibfield  {title} {\bibinfo {title} {Mixed-stacking few-layer
			graphene as an elemental weak ferroelectric material},\ }\href
	{https://doi.org/10.1021/acs.nanolett.2c04723} {\bibfield  {journal}
		{\bibinfo  {journal} {Nano Lett.}\ }\textbf {\bibinfo {volume} {23}},\
		\bibinfo {pages} {4120} (\bibinfo {year} {2023})}\BibitemShut {NoStop}%
	\bibitem [{\citenamefont {Yang}\ \emph {et~al.}(2023)\citenamefont {Yang},
		\citenamefont {Ding}, \citenamefont {Gao},\ and\ \citenamefont
		{Wu}}]{yang2023Atypical}%
	\BibitemOpen
	\bibfield  {author} {\bibinfo {author} {\bibfnamefont {L.}~\bibnamefont
			{Yang}}, \bibinfo {author} {\bibfnamefont {S.}~\bibnamefont {Ding}}, \bibinfo
		{author} {\bibfnamefont {J.}~\bibnamefont {Gao}},\ and\ \bibinfo {author}
		{\bibfnamefont {M.}~\bibnamefont {Wu}},\ }\bibfield  {title} {\bibinfo
		{title} {Atypical sliding and moir{\'e} ferroelectricity in pure multilayer
			graphene},\ }\href {https://doi.org/10.1103/PhysRevLett.131.096801}
	{\bibfield  {journal} {\bibinfo  {journal} {Phys. Rev. Lett.}\ }\textbf
		{\bibinfo {volume} {131}},\ \bibinfo {pages} {096801} (\bibinfo {year}
		{2023})}\BibitemShut {NoStop}%
	\bibitem [{\citenamefont {Vizner~Stern}\ \emph {et~al.}(2021)\citenamefont
		{Vizner~Stern}, \citenamefont {Waschitz}, \citenamefont {Cao}, \citenamefont
		{Nevo}, \citenamefont {Watanabe}, \citenamefont {Taniguchi}, \citenamefont
		{Sela}, \citenamefont {Urbakh}, \citenamefont {Hod},\ and\ \citenamefont
		{Ben~Shalom}}]{Stern2021interfacial}%
	\BibitemOpen
	\bibfield  {author} {\bibinfo {author} {\bibfnamefont {M.}~\bibnamefont
			{Vizner~Stern}}, \bibinfo {author} {\bibfnamefont {Y.}~\bibnamefont
			{Waschitz}}, \bibinfo {author} {\bibfnamefont {W.}~\bibnamefont {Cao}},
		\bibinfo {author} {\bibfnamefont {I.}~\bibnamefont {Nevo}}, \bibinfo {author}
		{\bibfnamefont {K.}~\bibnamefont {Watanabe}}, \bibinfo {author}
		{\bibfnamefont {T.}~\bibnamefont {Taniguchi}}, \bibinfo {author}
		{\bibfnamefont {E.}~\bibnamefont {Sela}}, \bibinfo {author} {\bibfnamefont
			{M.}~\bibnamefont {Urbakh}}, \bibinfo {author} {\bibfnamefont
			{O.}~\bibnamefont {Hod}},\ and\ \bibinfo {author} {\bibfnamefont
			{M.}~\bibnamefont {Ben~Shalom}},\ }\bibfield  {title} {\bibinfo {title}
		{Interfacial ferroelectricity by van der {Waals} sliding},\ }\href
	{https://doi.org/10.1126/science.abe8177} {\bibfield  {journal} {\bibinfo
			{journal} {Science}\ }\textbf {\bibinfo {volume} {372}},\ \bibinfo {pages}
		{1462} (\bibinfo {year} {2021})}\BibitemShut {NoStop}%
	\bibitem [{\citenamefont {Ko}\ \emph {et~al.}(2023)\citenamefont {Ko},
		\citenamefont {Yuk}, \citenamefont {Engelke}, \citenamefont {Carr},
		\citenamefont {Kim}, \citenamefont {Park}, \citenamefont {Heo}, \citenamefont
		{Kim}, \citenamefont {Kim}, \citenamefont {Kim}, \citenamefont {Taniguchi},
		\citenamefont {Watanabe}, \citenamefont {Park}, \citenamefont {Kaxiras},
		\citenamefont {Yang}, \citenamefont {Kim},\ and\ \citenamefont
		{Yoo}}]{ko2023Operando}%
	\BibitemOpen
	\bibfield  {author} {\bibinfo {author} {\bibfnamefont {K.}~\bibnamefont
			{Ko}}, \bibinfo {author} {\bibfnamefont {A.}~\bibnamefont {Yuk}}, \bibinfo
		{author} {\bibfnamefont {R.}~\bibnamefont {Engelke}}, \bibinfo {author}
		{\bibfnamefont {S.}~\bibnamefont {Carr}}, \bibinfo {author} {\bibfnamefont
			{J.}~\bibnamefont {Kim}}, \bibinfo {author} {\bibfnamefont {D.}~\bibnamefont
			{Park}}, \bibinfo {author} {\bibfnamefont {H.}~\bibnamefont {Heo}}, \bibinfo
		{author} {\bibfnamefont {H.-M.}\ \bibnamefont {Kim}}, \bibinfo {author}
		{\bibfnamefont {S.-G.}\ \bibnamefont {Kim}}, \bibinfo {author} {\bibfnamefont
			{H.}~\bibnamefont {Kim}}, \bibinfo {author} {\bibfnamefont {T.}~\bibnamefont
			{Taniguchi}}, \bibinfo {author} {\bibfnamefont {K.}~\bibnamefont {Watanabe}},
		\bibinfo {author} {\bibfnamefont {H.}~\bibnamefont {Park}}, \bibinfo {author}
		{\bibfnamefont {E.}~\bibnamefont {Kaxiras}}, \bibinfo {author} {\bibfnamefont
			{S.~M.}\ \bibnamefont {Yang}}, \bibinfo {author} {\bibfnamefont
			{P.}~\bibnamefont {Kim}},\ and\ \bibinfo {author} {\bibfnamefont
			{H.}~\bibnamefont {Yoo}},\ }\bibfield  {title} {\bibinfo {title} {Operando
			electron microscopy investigation of polar domain dynamics in twisted van der
			{{Waals}} homobilayers},\ }\href {https://doi.org/10.1038/s41563-023-01595-0}
	{\bibfield  {journal} {\bibinfo  {journal} {Nat. Mater.}\ }\textbf {\bibinfo
			{volume} {22}},\ \bibinfo {pages} {992} (\bibinfo {year} {2023})}\BibitemShut
	{NoStop}%
	\bibitem [{\citenamefont {Molino}\ \emph {et~al.}(2023)\citenamefont {Molino},
		\citenamefont {Aggarwal}, \citenamefont {Enaldiev}, \citenamefont
		{Plumadore}, \citenamefont {I.~Fal'ko},\ and\ \citenamefont
		{Luican-Mayer}}]{Molino2023Ferroelectric}%
	\BibitemOpen
	\bibfield  {author} {\bibinfo {author} {\bibfnamefont {L.}~\bibnamefont
			{Molino}}, \bibinfo {author} {\bibfnamefont {L.}~\bibnamefont {Aggarwal}},
		\bibinfo {author} {\bibfnamefont {V.}~\bibnamefont {Enaldiev}}, \bibinfo
		{author} {\bibfnamefont {R.}~\bibnamefont {Plumadore}}, \bibinfo {author}
		{\bibfnamefont {V.}~\bibnamefont {I.~Fal'ko}},\ and\ \bibinfo {author}
		{\bibfnamefont {A.}~\bibnamefont {Luican-Mayer}},\ }\bibfield  {title}
	{\bibinfo {title} {Ferroelectric switching at symmetry-broken interfaces by
			local control of dislocations networks},\ }\href
	{https://doi.org/https://doi.org/10.1002/adma.202207816} {\bibfield
		{journal} {\bibinfo  {journal} {Adv. Mater.}\ }\textbf {\bibinfo {volume}
			{35}},\ \bibinfo {pages} {2207816} (\bibinfo {year} {2023})}\BibitemShut
	{NoStop}%
	\bibitem [{\citenamefont {Zhang}\ \emph {et~al.}(2023)\citenamefont {Zhang},
		\citenamefont {Liu}, \citenamefont {Sun}, \citenamefont {Chen}, \citenamefont
		{Li}, \citenamefont {Moore}, \citenamefont {Liu}, \citenamefont {Wang},
		\citenamefont {Rossi}, \citenamefont {Jing}, \citenamefont {Fonseca},
		\citenamefont {Yang}, \citenamefont {Shao}, \citenamefont {Huang},
		\citenamefont {Handa}, \citenamefont {Xiong}, \citenamefont {Fu},
		\citenamefont {Pan}, \citenamefont {Halbertal}, \citenamefont {Xu},
		\citenamefont {Zheng}, \citenamefont {Schuck}, \citenamefont {Pasupathy},
		\citenamefont {Dean}, \citenamefont {Zhu}, \citenamefont {Cobden},
		\citenamefont {Xu}, \citenamefont {Liu}, \citenamefont {Fogler},
		\citenamefont {Hone},\ and\ \citenamefont {Basov}}]{Zhang2023Visualizing}%
	\BibitemOpen
	\bibfield  {author} {\bibinfo {author} {\bibfnamefont {S.}~\bibnamefont
			{Zhang}}, \bibinfo {author} {\bibfnamefont {Y.}~\bibnamefont {Liu}}, \bibinfo
		{author} {\bibfnamefont {Z.}~\bibnamefont {Sun}}, \bibinfo {author}
		{\bibfnamefont {X.}~\bibnamefont {Chen}}, \bibinfo {author} {\bibfnamefont
			{B.}~\bibnamefont {Li}}, \bibinfo {author} {\bibfnamefont {S.~L.}\
			\bibnamefont {Moore}}, \bibinfo {author} {\bibfnamefont {S.}~\bibnamefont
			{Liu}}, \bibinfo {author} {\bibfnamefont {Z.}~\bibnamefont {Wang}}, \bibinfo
		{author} {\bibfnamefont {S.~E.}\ \bibnamefont {Rossi}}, \bibinfo {author}
		{\bibfnamefont {R.}~\bibnamefont {Jing}}, \bibinfo {author} {\bibfnamefont
			{J.}~\bibnamefont {Fonseca}}, \bibinfo {author} {\bibfnamefont
			{B.}~\bibnamefont {Yang}}, \bibinfo {author} {\bibfnamefont {Y.}~\bibnamefont
			{Shao}}, \bibinfo {author} {\bibfnamefont {C.-Y.}\ \bibnamefont {Huang}},
		\bibinfo {author} {\bibfnamefont {T.}~\bibnamefont {Handa}}, \bibinfo
		{author} {\bibfnamefont {L.}~\bibnamefont {Xiong}}, \bibinfo {author}
		{\bibfnamefont {M.}~\bibnamefont {Fu}}, \bibinfo {author} {\bibfnamefont
			{T.-C.}\ \bibnamefont {Pan}}, \bibinfo {author} {\bibfnamefont
			{D.}~\bibnamefont {Halbertal}}, \bibinfo {author} {\bibfnamefont
			{X.}~\bibnamefont {Xu}}, \bibinfo {author} {\bibfnamefont {W.}~\bibnamefont
			{Zheng}}, \bibinfo {author} {\bibfnamefont {P.~J.}\ \bibnamefont {Schuck}},
		\bibinfo {author} {\bibfnamefont {A.~N.}\ \bibnamefont {Pasupathy}}, \bibinfo
		{author} {\bibfnamefont {C.~R.}\ \bibnamefont {Dean}}, \bibinfo {author}
		{\bibfnamefont {X.}~\bibnamefont {Zhu}}, \bibinfo {author} {\bibfnamefont
			{D.~H.}\ \bibnamefont {Cobden}}, \bibinfo {author} {\bibfnamefont
			{X.}~\bibnamefont {Xu}}, \bibinfo {author} {\bibfnamefont {M.}~\bibnamefont
			{Liu}}, \bibinfo {author} {\bibfnamefont {M.~M.}\ \bibnamefont {Fogler}},
		\bibinfo {author} {\bibfnamefont {J.~C.}\ \bibnamefont {Hone}},\ and\
		\bibinfo {author} {\bibfnamefont {D.~N.}\ \bibnamefont {Basov}},\ }\bibfield
	{title} {\bibinfo {title} {Visualizing moir{\'e} ferroelectricity via
			plasmons and nano-photocurrent in graphene/twisted-{WS}e$_2$ structures},\
	}\href@noop {} {\bibfield  {journal} {\bibinfo  {journal} {Nat. Commun.}\
		}\textbf {\bibinfo {volume} {14}},\ \bibinfo {pages} {6200} (\bibinfo {year}
		{2023})}\BibitemShut {NoStop}%
	\bibitem [{\citenamefont {Bousquet}\ \emph {et~al.}(2011)\citenamefont
		{Bousquet}, \citenamefont {Spaldin},\ and\ \citenamefont
		{Delaney}}]{Bousquet2011Unexpectedly}%
	\BibitemOpen
	\bibfield  {author} {\bibinfo {author} {\bibfnamefont {E.}~\bibnamefont
			{Bousquet}}, \bibinfo {author} {\bibfnamefont {N.~A.}\ \bibnamefont
			{Spaldin}},\ and\ \bibinfo {author} {\bibfnamefont {K.~T.}\ \bibnamefont
			{Delaney}},\ }\bibfield  {title} {\bibinfo {title} {Unexpectedly large
			electronic contribution to linear magnetoelectricity},\ }\href
	{https://doi.org/10.1103/PhysRevLett.106.107202} {\bibfield  {journal}
		{\bibinfo  {journal} {Phys. Rev. Lett.}\ }\textbf {\bibinfo {volume} {106}},\
		\bibinfo {pages} {107202} (\bibinfo {year} {2011})}\BibitemShut {NoStop}%
	\bibitem [{\citenamefont {Dasa}\ \emph {et~al.}(2019)\citenamefont {Dasa},
		\citenamefont {Hao}, \citenamefont {Liu},\ and\ \citenamefont
		{Xu}}]{Dasa2019Designing}%
	\BibitemOpen
	\bibfield  {author} {\bibinfo {author} {\bibfnamefont {T.~R.}\ \bibnamefont
			{Dasa}}, \bibinfo {author} {\bibfnamefont {L.}~\bibnamefont {Hao}}, \bibinfo
		{author} {\bibfnamefont {J.}~\bibnamefont {Liu}},\ and\ \bibinfo {author}
		{\bibfnamefont {H.}~\bibnamefont {Xu}},\ }\bibfield  {title} {\bibinfo
		{title} {Designing iridate-based superlattice with large magnetoelectric
			coupling},\ }\href@noop {} {\bibfield  {journal} {\bibinfo  {journal} {J.
				Mater. Chem. C}\ }\textbf {\bibinfo {volume} {7}},\ \bibinfo {pages} {13294}
		(\bibinfo {year} {2019})}\BibitemShut {NoStop}%
	\bibitem [{\citenamefont {Feng}\ \emph {et~al.}(2019)\citenamefont {Feng},
		\citenamefont {Jiang}, \citenamefont {Wu}, \citenamefont {Li}, \citenamefont
		{He}, \citenamefont {Ma}, \citenamefont {Xue},\ and\ \citenamefont
		{Wang}}]{Feng2019Tunable}%
	\BibitemOpen
	\bibfield  {author} {\bibinfo {author} {\bibfnamefont {Y.}~\bibnamefont
			{Feng}}, \bibinfo {author} {\bibfnamefont {G.}~\bibnamefont {Jiang}},
		\bibinfo {author} {\bibfnamefont {W.}~\bibnamefont {Wu}}, \bibinfo {author}
		{\bibfnamefont {S.}~\bibnamefont {Li}}, \bibinfo {author} {\bibfnamefont
			{K.}~\bibnamefont {He}}, \bibinfo {author} {\bibfnamefont {X.}~\bibnamefont
			{Ma}}, \bibinfo {author} {\bibfnamefont {Q.-K.}\ \bibnamefont {Xue}},\ and\
		\bibinfo {author} {\bibfnamefont {Y.}~\bibnamefont {Wang}},\ }\bibfield
	{title} {\bibinfo {title} {Tunable chiral and helical edge state transport in
			a magnetic topological insulator bilayer},\ }\href
	{https://doi.org/10.1103/PhysRevB.100.165403} {\bibfield  {journal} {\bibinfo
			{journal} {Phys. Rev. B}\ }\textbf {\bibinfo {volume} {100}},\ \bibinfo
		{pages} {165403} (\bibinfo {year} {2019})}\BibitemShut {NoStop}%
	\bibitem [{\citenamefont {Rosen}\ \emph {et~al.}(2017)\citenamefont {Rosen},
		\citenamefont {Fox}, \citenamefont {Kou}, \citenamefont {Pan}, \citenamefont
		{Wang},\ and\ \citenamefont {Goldhaber-Gordon}}]{Rosen2017Chiral}%
	\BibitemOpen
	\bibfield  {author} {\bibinfo {author} {\bibfnamefont {I.~T.}\ \bibnamefont
			{Rosen}}, \bibinfo {author} {\bibfnamefont {E.~J.}\ \bibnamefont {Fox}},
		\bibinfo {author} {\bibfnamefont {X.}~\bibnamefont {Kou}}, \bibinfo {author}
		{\bibfnamefont {L.}~\bibnamefont {Pan}}, \bibinfo {author} {\bibfnamefont
			{K.~L.}\ \bibnamefont {Wang}},\ and\ \bibinfo {author} {\bibfnamefont
			{D.}~\bibnamefont {Goldhaber-Gordon}},\ }\bibfield  {title} {\bibinfo {title}
		{Chiral transport along magnetic domain walls in the quantum anomalous {Hall}
			effect},\ }\href {https://doi.org/10.1038/s41535-017-0073-0} {\bibfield
		{journal} {\bibinfo  {journal} {npj Quant. Mater.}\ }\textbf {\bibinfo
			{volume} {2}},\ \bibinfo {pages} {69} (\bibinfo {year} {2017})}\BibitemShut
	{NoStop}%
	\bibitem [{\citenamefont {Zhu}\ \emph {et~al.}(2025)\citenamefont {Zhu},
		\citenamefont {Feng}, \citenamefont {Zhou}, \citenamefont {Wang},
		\citenamefont {Yao}, \citenamefont {Lian}, \citenamefont {Lin}, \citenamefont
		{He}, \citenamefont {Lin}, \citenamefont {Wang}, \citenamefont {Wang},
		\citenamefont {Yang}, \citenamefont {Li}, \citenamefont {Wu}, \citenamefont
		{Liu}, \citenamefont {Wang}, \citenamefont {Shen}, \citenamefont {Zhang},
		\citenamefont {Wang},\ and\ \citenamefont {Wang}}]{Zhu2025Direct}%
	\BibitemOpen
	\bibfield  {author} {\bibinfo {author} {\bibfnamefont {J.}~\bibnamefont
			{Zhu}}, \bibinfo {author} {\bibfnamefont {Y.}~\bibnamefont {Feng}}, \bibinfo
		{author} {\bibfnamefont {X.}~\bibnamefont {Zhou}}, \bibinfo {author}
		{\bibfnamefont {Y.}~\bibnamefont {Wang}}, \bibinfo {author} {\bibfnamefont
			{H.}~\bibnamefont {Yao}}, \bibinfo {author} {\bibfnamefont {Z.}~\bibnamefont
			{Lian}}, \bibinfo {author} {\bibfnamefont {W.}~\bibnamefont {Lin}}, \bibinfo
		{author} {\bibfnamefont {Q.}~\bibnamefont {He}}, \bibinfo {author}
		{\bibfnamefont {Y.}~\bibnamefont {Lin}}, \bibinfo {author} {\bibfnamefont
			{Y.}~\bibnamefont {Wang}}, \bibinfo {author} {\bibfnamefont {Y.}~\bibnamefont
			{Wang}}, \bibinfo {author} {\bibfnamefont {S.}~\bibnamefont {Yang}}, \bibinfo
		{author} {\bibfnamefont {H.}~\bibnamefont {Li}}, \bibinfo {author}
		{\bibfnamefont {Y.}~\bibnamefont {Wu}}, \bibinfo {author} {\bibfnamefont
			{C.}~\bibnamefont {Liu}}, \bibinfo {author} {\bibfnamefont {J.}~\bibnamefont
			{Wang}}, \bibinfo {author} {\bibfnamefont {J.}~\bibnamefont {Shen}}, \bibinfo
		{author} {\bibfnamefont {J.}~\bibnamefont {Zhang}}, \bibinfo {author}
		{\bibfnamefont {Y.}~\bibnamefont {Wang}},\ and\ \bibinfo {author}
		{\bibfnamefont {Y.}~\bibnamefont {Wang}},\ }\bibfield  {title} {\bibinfo
		{title} {Direct observation of chiral edge current at zero magnetic field in
			a magnetic topological insulator},\ }\href@noop {} {\bibfield  {journal}
		{\bibinfo  {journal} {Nat. Commun.}\ }\textbf {\bibinfo {volume} {16}},\
		\bibinfo {pages} {963} (\bibinfo {year} {2025})}\BibitemShut {NoStop}%
	\bibitem [{\citenamefont {Saito}\ \emph {et~al.}(2020)\citenamefont {Saito},
		\citenamefont {Ge}, \citenamefont {Watanabe}, \citenamefont {Taniguchi},\
		and\ \citenamefont {Young}}]{Saito2020Independent}%
	\BibitemOpen
	\bibfield  {author} {\bibinfo {author} {\bibfnamefont {Y.}~\bibnamefont
			{Saito}}, \bibinfo {author} {\bibfnamefont {J.}~\bibnamefont {Ge}}, \bibinfo
		{author} {\bibfnamefont {K.}~\bibnamefont {Watanabe}}, \bibinfo {author}
		{\bibfnamefont {T.}~\bibnamefont {Taniguchi}},\ and\ \bibinfo {author}
		{\bibfnamefont {A.~F.}\ \bibnamefont {Young}},\ }\bibfield  {title} {\bibinfo
		{title} {Independent superconductors and correlated insulators in twisted
			bilayer graphene},\ }\href {https://doi.org/10.1038/s41567-020-0928-3}
	{\bibfield  {journal} {\bibinfo  {journal} {Nat. Phys.}\ }\textbf {\bibinfo
			{volume} {16}},\ \bibinfo {pages} {926} (\bibinfo {year} {2020})}\BibitemShut
	{NoStop}%
	\bibitem [{\citenamefont {You}\ \emph {et~al.}(2008)\citenamefont {You},
		\citenamefont {Ni}, \citenamefont {Yu},\ and\ \citenamefont
		{Shen}}]{You2008Edge}%
	\BibitemOpen
	\bibfield  {author} {\bibinfo {author} {\bibfnamefont {Y.}~\bibnamefont
			{You}}, \bibinfo {author} {\bibfnamefont {Z.}~\bibnamefont {Ni}}, \bibinfo
		{author} {\bibfnamefont {T.}~\bibnamefont {Yu}},\ and\ \bibinfo {author}
		{\bibfnamefont {Z.}~\bibnamefont {Shen}},\ }\bibfield  {title} {\bibinfo
		{title} {Edge chirality determination of graphene by {R}aman spectroscopy},\
	}\href {https://doi.org/10.1063/1.3005599} {\bibfield  {journal} {\bibinfo
			{journal} {Appl. Phys. Lett.}\ }\textbf {\bibinfo {volume} {93}},\ \bibinfo
		{pages} {163112} (\bibinfo {year} {2008})}\BibitemShut {NoStop}%
	\bibitem [{\citenamefont {Wang}\ \emph {et~al.}(2013)\citenamefont {Wang},
		\citenamefont {Meric}, \citenamefont {Huang}, \citenamefont {Gao},
		\citenamefont {Gao}, \citenamefont {Tran}, \citenamefont {Taniguchi},
		\citenamefont {Watanabe}, \citenamefont {Campos}, \citenamefont {Muller},
		\citenamefont {Guo}, \citenamefont {Kim}, \citenamefont {Hone}, \citenamefont
		{Shepard},\ and\ \citenamefont {Dean}}]{Wang2013one}%
	\BibitemOpen
	\bibfield  {author} {\bibinfo {author} {\bibfnamefont {L.}~\bibnamefont
			{Wang}}, \bibinfo {author} {\bibfnamefont {I.}~\bibnamefont {Meric}},
		\bibinfo {author} {\bibfnamefont {P.~Y.}\ \bibnamefont {Huang}}, \bibinfo
		{author} {\bibfnamefont {Q.}~\bibnamefont {Gao}}, \bibinfo {author}
		{\bibfnamefont {Y.}~\bibnamefont {Gao}}, \bibinfo {author} {\bibfnamefont
			{H.}~\bibnamefont {Tran}}, \bibinfo {author} {\bibfnamefont {T.}~\bibnamefont
			{Taniguchi}}, \bibinfo {author} {\bibfnamefont {K.}~\bibnamefont {Watanabe}},
		\bibinfo {author} {\bibfnamefont {L.~M.}\ \bibnamefont {Campos}}, \bibinfo
		{author} {\bibfnamefont {D.~A.}\ \bibnamefont {Muller}}, \bibinfo {author}
		{\bibfnamefont {J.}~\bibnamefont {Guo}}, \bibinfo {author} {\bibfnamefont
			{P.}~\bibnamefont {Kim}}, \bibinfo {author} {\bibfnamefont {J.}~\bibnamefont
			{Hone}}, \bibinfo {author} {\bibfnamefont {K.~L.}\ \bibnamefont {Shepard}},\
		and\ \bibinfo {author} {\bibfnamefont {C.~R.}\ \bibnamefont {Dean}},\
	}\bibfield  {title} {\bibinfo {title} {One-dimensional electrical contact to
			a two-dimensional material},\ }\href
	{https://doi.org/10.1126/science.1244358} {\bibfield  {journal} {\bibinfo
			{journal} {Science}\ }\textbf {\bibinfo {volume} {342}},\ \bibinfo {pages}
		{614} (\bibinfo {year} {2013})}\BibitemShut {NoStop}%
	\bibitem [{\citenamefont {Kim}\ \emph {et~al.}(2016)\citenamefont {Kim},
		\citenamefont {Yankowitz}, \citenamefont {Fallahazad}, \citenamefont {Kang},
		\citenamefont {Movva}, \citenamefont {Huang}, \citenamefont {Larentis},
		\citenamefont {Corbet}, \citenamefont {Taniguchi}, \citenamefont {Watanabe},
		\citenamefont {Banerjee}, \citenamefont {LeRoy},\ and\ \citenamefont
		{Tutuc}}]{Kim2016van}%
	\BibitemOpen
	\bibfield  {author} {\bibinfo {author} {\bibfnamefont {K.}~\bibnamefont
			{Kim}}, \bibinfo {author} {\bibfnamefont {M.}~\bibnamefont {Yankowitz}},
		\bibinfo {author} {\bibfnamefont {B.}~\bibnamefont {Fallahazad}}, \bibinfo
		{author} {\bibfnamefont {S.}~\bibnamefont {Kang}}, \bibinfo {author}
		{\bibfnamefont {H.~C.~P.}\ \bibnamefont {Movva}}, \bibinfo {author}
		{\bibfnamefont {S.}~\bibnamefont {Huang}}, \bibinfo {author} {\bibfnamefont
			{S.}~\bibnamefont {Larentis}}, \bibinfo {author} {\bibfnamefont {C.~M.}\
			\bibnamefont {Corbet}}, \bibinfo {author} {\bibfnamefont {T.}~\bibnamefont
			{Taniguchi}}, \bibinfo {author} {\bibfnamefont {K.}~\bibnamefont {Watanabe}},
		\bibinfo {author} {\bibfnamefont {S.~K.}\ \bibnamefont {Banerjee}}, \bibinfo
		{author} {\bibfnamefont {B.~J.}\ \bibnamefont {LeRoy}},\ and\ \bibinfo
		{author} {\bibfnamefont {E.}~\bibnamefont {Tutuc}},\ }\bibfield  {title}
	{\bibinfo {title} {van der {W}aals heterostructures with high accuracy
			rotational alignment},\ }\href {https://doi.org/10.1021/acs.nanolett.5b05263}
	{\bibfield  {journal} {\bibinfo  {journal} {Nano Lett.}\ }\textbf {\bibinfo
			{volume} {16}},\ \bibinfo {pages} {1989} (\bibinfo {year}
		{2016})}\BibitemShut {NoStop}%
	\bibitem [{\citenamefont {Cao}\ \emph {et~al.}(2016)\citenamefont {Cao},
		\citenamefont {Luo}, \citenamefont {Fatemi}, \citenamefont {Fang},
		\citenamefont {Sanchez-Yamagishi}, \citenamefont {Watanabe}, \citenamefont
		{Taniguchi}, \citenamefont {Kaxiras},\ and\ \citenamefont
		{Jarillo-Herrero}}]{cao2016superlattice}%
	\BibitemOpen
	\bibfield  {author} {\bibinfo {author} {\bibfnamefont {Y.}~\bibnamefont
			{Cao}}, \bibinfo {author} {\bibfnamefont {J.~Y.}\ \bibnamefont {Luo}},
		\bibinfo {author} {\bibfnamefont {V.}~\bibnamefont {Fatemi}}, \bibinfo
		{author} {\bibfnamefont {S.}~\bibnamefont {Fang}}, \bibinfo {author}
		{\bibfnamefont {J.~D.}\ \bibnamefont {Sanchez-Yamagishi}}, \bibinfo {author}
		{\bibfnamefont {K.}~\bibnamefont {Watanabe}}, \bibinfo {author}
		{\bibfnamefont {T.}~\bibnamefont {Taniguchi}}, \bibinfo {author}
		{\bibfnamefont {E.}~\bibnamefont {Kaxiras}},\ and\ \bibinfo {author}
		{\bibfnamefont {P.}~\bibnamefont {Jarillo-Herrero}},\ }\bibfield  {title}
	{\bibinfo {title} {Superlattice-induced insulating states and
			valley-protected orbits in twisted bilayer graphene},\ }\href
	{https://doi.org/10.1103/PhysRevLett.117.116804} {\bibfield  {journal}
		{\bibinfo  {journal} {Phys. Rev. Lett.}\ }\textbf {\bibinfo {volume} {117}},\
		\bibinfo {pages} {116804} (\bibinfo {year} {2016})}\BibitemShut {NoStop}%
	\bibitem [{\citenamefont {Kresse}\ and\ \citenamefont
		{Furthmüller}(1996)}]{KRESSE1996Efficiency}%
	\BibitemOpen
	\bibfield  {author} {\bibinfo {author} {\bibfnamefont {G.}~\bibnamefont
			{Kresse}}\ and\ \bibinfo {author} {\bibfnamefont {J.}~\bibnamefont
			{Furthmüller}},\ }\bibfield  {title} {\bibinfo {title} {Efficiency of
			ab-initio total energy calculations for metals and semiconductors using a
			plane-wave basis set},\ }\href
	{https://doi.org/https://doi.org/10.1016/0927-0256(96)00008-0} {\bibfield
		{journal} {\bibinfo  {journal} {Comput. Mater. Sci.}\ }\textbf {\bibinfo
			{volume} {6}},\ \bibinfo {pages} {15} (\bibinfo {year} {1996})}\BibitemShut
	{NoStop}%
	\bibitem [{\citenamefont {Kresse}\ and\ \citenamefont
		{Furthm\"uller}(1996)}]{Kresse1996Efficient}%
	\BibitemOpen
	\bibfield  {author} {\bibinfo {author} {\bibfnamefont {G.}~\bibnamefont
			{Kresse}}\ and\ \bibinfo {author} {\bibfnamefont {J.}~\bibnamefont
			{Furthm\"uller}},\ }\bibfield  {title} {\bibinfo {title} {Efficient iterative
			schemes for ab initio total-energy calculations using a plane-wave basis
			set},\ }\href {https://doi.org/10.1103/PhysRevB.54.11169} {\bibfield
		{journal} {\bibinfo  {journal} {Phys. Rev. B}\ }\textbf {\bibinfo {volume}
			{54}},\ \bibinfo {pages} {11169} (\bibinfo {year} {1996})}\BibitemShut
	{NoStop}%
	\bibitem [{\citenamefont {Perdew}\ \emph {et~al.}(1996)\citenamefont {Perdew},
		\citenamefont {Burke},\ and\ \citenamefont
		{Ernzerhof}}]{Perdew1996Generalized}%
	\BibitemOpen
	\bibfield  {author} {\bibinfo {author} {\bibfnamefont {J.~P.}\ \bibnamefont
			{Perdew}}, \bibinfo {author} {\bibfnamefont {K.}~\bibnamefont {Burke}},\ and\
		\bibinfo {author} {\bibfnamefont {M.}~\bibnamefont {Ernzerhof}},\ }\bibfield
	{title} {\bibinfo {title} {Generalized gradient approximation made simple},\
	}\href {https://doi.org/10.1103/PhysRevLett.77.3865} {\bibfield  {journal}
		{\bibinfo  {journal} {Phys. Rev. Lett.}\ }\textbf {\bibinfo {volume} {77}},\
		\bibinfo {pages} {3865} (\bibinfo {year} {1996})}\BibitemShut {NoStop}%
	\bibitem [{\citenamefont {Klime\ifmmode\check{s}\else\v{s}\fi{}}\ \emph
		{et~al.}(2011)\citenamefont {Klime\ifmmode\check{s}\else\v{s}\fi{}},
		\citenamefont {Bowler},\ and\ \citenamefont {Michaelides}}]{Klime2011Van}%
	\BibitemOpen
	\bibfield  {author} {\bibinfo {author} {\bibfnamefont {J.}~\bibnamefont
			{Klime\ifmmode\check{s}\else\v{s}\fi{}}}, \bibinfo {author} {\bibfnamefont
			{D.~R.}\ \bibnamefont {Bowler}},\ and\ \bibinfo {author} {\bibfnamefont
			{A.}~\bibnamefont {Michaelides}},\ }\bibfield  {title} {\bibinfo {title} {Van
			der {W}aals density functionals applied to solids},\ }\href
	{https://doi.org/10.1103/PhysRevB.83.195131} {\bibfield  {journal} {\bibinfo
			{journal} {Phys. Rev. B}\ }\textbf {\bibinfo {volume} {83}},\ \bibinfo
		{pages} {195131} (\bibinfo {year} {2011})}\BibitemShut {NoStop}%
	\bibitem [{\citenamefont {Neugebauer}\ and\ \citenamefont
		{Scheffler}(1992)}]{Neugebauer1992Adsorbate}%
	\BibitemOpen
	\bibfield  {author} {\bibinfo {author} {\bibfnamefont {J.}~\bibnamefont
			{Neugebauer}}\ and\ \bibinfo {author} {\bibfnamefont {M.}~\bibnamefont
			{Scheffler}},\ }\bibfield  {title} {\bibinfo {title} {Adsorbate-substrate and
			adsorbate-adsorbate interactions of {Na} and {K} adlayers on {Al(111)}},\
	}\href {https://doi.org/10.1103/PhysRevB.46.16067} {\bibfield  {journal}
		{\bibinfo  {journal} {Phys. Rev. B}\ }\textbf {\bibinfo {volume} {46}},\
		\bibinfo {pages} {16067} (\bibinfo {year} {1992})}\BibitemShut {NoStop}%
	\bibitem [{\citenamefont {Henkelman}\ \emph {et~al.}(2000)\citenamefont
		{Henkelman}, \citenamefont {Uberuaga},\ and\ \citenamefont
		{J\'{o}nsson}}]{Henkelman2000climbing}%
	\BibitemOpen
	\bibfield  {author} {\bibinfo {author} {\bibfnamefont {G.}~\bibnamefont
			{Henkelman}}, \bibinfo {author} {\bibfnamefont {B.~P.}\ \bibnamefont
			{Uberuaga}},\ and\ \bibinfo {author} {\bibfnamefont {H.}~\bibnamefont
			{J\'{o}nsson}},\ }\bibfield  {title} {\bibinfo {title} {A climbing image
			nudged elastic band method for finding saddle points and minimum energy
			paths},\ }\href {https://doi.org/10.1063/1.1329672} {\bibfield  {journal}
		{\bibinfo  {journal} {J. Chem. Phys.}\ }\textbf {\bibinfo {volume} {113}},\
		\bibinfo {pages} {9901} (\bibinfo {year} {2000})}\BibitemShut {NoStop}%
	\bibitem [{\citenamefont {King-Smith}\ and\ \citenamefont
		{Vanderbilt}(1993)}]{King-Smith1993Theory}%
	\BibitemOpen
	\bibfield  {author} {\bibinfo {author} {\bibfnamefont {R.~D.}\ \bibnamefont
			{King-Smith}}\ and\ \bibinfo {author} {\bibfnamefont {D.}~\bibnamefont
			{Vanderbilt}},\ }\bibfield  {title} {\bibinfo {title} {Theory of polarization
			of crystalline solids},\ }\href {https://doi.org/10.1103/PhysRevB.47.1651}
	{\bibfield  {journal} {\bibinfo  {journal} {Phys. Rev. B}\ }\textbf {\bibinfo
			{volume} {47}},\ \bibinfo {pages} {1651} (\bibinfo {year}
		{1993})}\BibitemShut {NoStop}%
	\bibitem [{\citenamefont {Gonze}\ and\ \citenamefont
		{Lee}(1997)}]{Gonze1997Dynamical}%
	\BibitemOpen
	\bibfield  {author} {\bibinfo {author} {\bibfnamefont {X.}~\bibnamefont
			{Gonze}}\ and\ \bibinfo {author} {\bibfnamefont {C.}~\bibnamefont {Lee}},\
	}\bibfield  {title} {\bibinfo {title} {Dynamical matrices, {B}orn effective
			charges, dielectric permittivity tensors, and interatomic force constants
			from density-functional perturbation theory},\ }\href
	{https://doi.org/10.1103/PhysRevB.55.10355} {\bibfield  {journal} {\bibinfo
			{journal} {Phys. Rev. B}\ }\textbf {\bibinfo {volume} {55}},\ \bibinfo
		{pages} {10355} (\bibinfo {year} {1997})}\BibitemShut {NoStop}%
\end{thebibliography}
\end{document}